\documentclass[journal]{IEEEtran}

\usepackage{subfigure}
\usepackage{amsmath,amssymb}
\usepackage[dvips]{graphicx}
\usepackage{amsfonts}
\usepackage[mathscr]{eucal}
\usepackage{latexsym}
\usepackage{amsthm}
\usepackage{exscale}
\usepackage[mathscr]{eucal}
\usepackage{bm}
\usepackage[dvipsnames]{color}
\usepackage{cases}
\usepackage{epsfig}
\usepackage[center,small]{caption}
\usepackage{algorithm}
\usepackage{algorithmic}
\usepackage[verbose,nospace,sort]{cite}
\usepackage{tabularx}
\usepackage{multirow}
\usepackage{multicol}
\usepackage{balance}

\graphicspath{{./figsJSAC/}}

\scrollmode

\newcommand{\RF}{\mathrm{RF}}
\newcommand{\CB}{\mathrm{CB}}
\newcommand{\MS}{\mathrm{MS}}
\newcommand{\BS}{\mathrm{BS}}

\newcommand{\D}{\tilde{D}}
\newcommand{\sinc}{\mathrm{sinc}}
\newcommand{\tr}{\mathrm{tr}}

\newtheorem{lemma}{Lemma}

\begin{document}
\title{Multi-User Millimeter Wave MIMO with Full-Dimensional Lens Antenna Array}
\author{Yong~Zeng, Lu~Yang, and Rui~Zhang\\
\thanks{The authors are with the Department of Electrical and Computer Engineering, National University of Singapore (e-mail: \{elezeng, eleylu, elezhang\}@nus.edu.sg).}
\thanks{Part of this paper has been submitted to the IEEE International Conference on Communications (ICC), May 21-25,  2017, Paris, France.}
\vspace{-3ex}}

\maketitle

\begin{abstract}
Millimeter wave (mmWave) communication by utilizing lens antenna arrays is a promising technique for realizing cost-effective 5G wireless systems with large MIMO (multiple-input multiple-output) but only limited radio frequency (RF) chains. This paper studies an uplink multi-user mmWave single-sided lens MIMO system, where only the base station (BS) is equipped with a full-dimensional (FD) lens antenna array with both elevation and azimuth angle resolution capabilities, and each mobile station (MS) employs the conventional uniform planar array (UPA) without the lens. By exploiting  the \emph{angle-dependent energy focusing} property of the lens antenna array at the BS as well as the \emph{multi-path sparsity} of mmWave channels,  we propose a low-complexity  \emph{path division multiple access} (PDMA) scheme, which enables virtually interference-free multi-user communications when the angle of arrivals (AoAs) of all MS multi-path signals are sufficiently separable at the BS. To this end, a new technique called \emph{path delay compensation} is proposed at the BS to effectively transform the multi-user frequency-selective MIMO channels to parallel frequency-flat small-size MIMO channels for different MSs, for each of which the low-complexity single-carrier (SC) transmission is applied. For general scenarios with insufficient AoA separations, analog beamforming at the MSs and digital combining at the BS are jointly designed to maximize the achievable sum-rate of the MSs based on their effective MIMO channels resulting from path delay compensation. In addition, we propose a new and efficient channel estimation scheme tailored for PDMA, which requires negligible training overhead in practical mmWave systems and yet leads to comparable performance as that based on perfect channel state information (CSI). Numerical results show that the proposed design significantly outperforms state-of-the-art benchmark systems in terms of sum-rate, but with significantly reduced hardware cost and signal processing complexity.
\end{abstract}

\begin{keywords}
Millimeter wave communication, MIMO, lens antenna array, path division multiple access (PDMA), analog beamforming, channel estimation.
\end{keywords}

\section{Introduction}
Due to the large  bandwidth available, millimeter wave (mmWave) communication over the spectrum above 28GHz has emerged as a key enabling technology for the fifth-generation (5G) wireless systems \cite{566,567,569,594,856}. In fact, mmWave systems at the unlicensed 60GHz have already been standardized for short-range applications, such as wireless personal area networks (WPAN) \cite{571} and wireless local area networks (WLAN) \cite{581}. In July 2016, the FCC (Federal Communications Commission) of the United States has released 3.85GHz of licensed and 7GHz of unlicensed mmWave frequency spectrum for future broadband communications \cite{850}, which made an important step forward for mmWave technology to be practically employed for mobile access in 5G.

Compared to the conventional sub-6GHz wireless systems, mmWave communications are faced with many new design challenges \cite{486}. In particular, to maintain a sufficient link margin for  signal coverage at a reasonable range, mmWave systems  typically require large antenna arrays (in terms of the number of array elements) to be equipped at the base station (BS), and also possibly at each mobile station (MS), to achieve highly directional communications. While packing more antennas compactly with small form factors is becoming more feasible at mmWave frequencies, thanks to the significantly reduced wavelength, the associated signal processing complexity and radio frequency (RF) chain cost in terms of hardware and power consumption increase dramatically, if the conventional multiple-input multiple-output (MIMO) communication techniques are used. 
Furthermore, due to the large signal bandwidth, mmWave systems are most likely to operate over frequency-selective channels, for which the detrimental inter-symbol interference (ISI) needs to be effectively mitigated. However, ISI mitigation is a non-trivial task for wide-band mmWave large MIMO systems, especially for practical designs that need to operate with low signal processing complexity and RF chain costs, due to which the traditional ISI mitigation techniques such as MIMO-OFDM (orthogonal frequency division multiplexing) and sophisticated time/frequency-domain equalizations become less effective.

Extensive research efforts have been devoted to developing various cost-aware mmWave communication techniques, such as analog beamforming \cite{574,575,573}, hybrid analog/digital processing \cite{576,577,578,592,825,827,579,832}, and low-resolution ADCs (analog to digital converters) for signal reception \cite{828}. In particular, hybrid analog/digital processing is regarded as an effective technique to offer flexible trade-offs between performance and cost by varying the number of RF chains at the transmitter/receiver. With this technique, the signal processing is implemented via two stages: a low-dimensional baseband digital processing using limited number of RF chains concatenated with an RF-band analog processing through a network of phase shifters. In \cite{578}, the hybrid precoding and combining matrices are optimized for single-user narrow-band mmWave systems based on the concept of orthogonal matching pursuit, and its extensions to codebook-based multi-user frequency-flat and single-user frequency-selective MIMO-OFDM  systems  are studied in \cite{592} and \cite{825}, respectively. In \cite{827}, it has been shown that for narrow-band systems, hybrid processing is able to achieve the same performance as the fully digital processing scheme, provided that the number of RF chains is at least twice of the number of data streams. Channel estimation for mmWave hybrid systems has also been investigated in \cite{579} and \cite{832}. However, one major drawback of the hybrid analog/digital architecture is the general requirement of an extremely large number of phase shifters, which consume extra power that may even outweigh the power saving due to the use of fewer RF chains. Besides, due to the additional constant-amplitude constraints on the analog beamformers as well as the limited RF chains available, the precoding/combining design as well as channel estimation for hybrid processing  is more complicated than the conventional fully digital design, especially for wide-band systems with frequency-selective channels \cite{825}, \cite{832}.     

Another line of researches has focused on achieving cost-effective mmWave communications by utilizing the advanced lens antenna arrays \cite{857,553,485,823,627,830,851,867}. A lens antenna array is in general composed of an electromagnetic (EM) lens with energy focusing capability, and a matching antenna array with elements located in the focal region of the lens. 
 In \cite{823}, it is derived that the array response of a lens antenna array can be expressed as a ``sinc''-type function in terms of the angle of arrival/departure (AoA/AoD) and the antenna locations, which theoretically confirms  the {\it angle-dependent energy focusing} property of lens antenna arrays, i.e., for each uniform plane wave of a given AoA/AoD, only those antennas located in the close vicinity of the energy focusing point receives/steers significant power. By exploiting this unique array response, together with the {\it multi-path sparsity} of mmWave channels \cite{569}, a new spatial multiplexing technique, termed {\it path division multiplexing (PDM)}, is proposed in \cite{823} for the point-to-point mmWave MIMO channel, which is capacity-achieving using the simple single-carrier (SC) transmission with low RF chain cost and signal processing complexity, even for wide-band frequency-selective mmWave communications. MmWave lens MIMO systems have also been shown be able to effectively reduce the channel estimation overhead as compared to hybrid processing systems subject to limited number of RF chains \cite{830}.

However, most existing works on mmWave lens MIMO consider only the single-user systems with double-sided lenses, i.e., by assuming lens antenna arrays at both the BS and MS. Due to the additional gap required between the EM lens and antenna elements, lens antenna array is expected to be more suitable to be employed at the BS than at the MS, which has more stringent size limitations in practice. Besides, existing lens array designs mostly ignore the signal's elevation angles, and thus consider only the two-dimensional (2D) array with antenna elements located in the focal arc only. In this paper, we consider a more general setup of multi-user mmWave lens MIMO system with single-sided full-dimensional (FD) lens antenna array, i.e., only the BS is equipped with an FD lens antenna array with both azimuth and elevation angle resolution capabilities, while each MS employs the conventional antenna array such as the uniform planar array (UPA) without the lens. Under such a new setup and with limited RF chains at the BS and one single RF chain at each MS, we propose a low-complexity transceiver design for general multi-user wide-band mmWave communications in frequency-selective channels, as well as an efficient channel estimation scheme with nearly negligible training overhead. The specific contributions of this paper are summarized as follows.
\begin{itemize}
\item First, we introduce the general architecture for FD lens antenna arrays. Different from the 2D lens array studied  in  our prior work \cite{823}, the array elements of FD lens array are located in the focal surface (instead of focal arc) of the lens and thus offer both elevation and azimuth angle resolution capabilities. We further derive the array response vector of the FD lens antenna array, which is shown to be given by the product of two ``sinc'' functions reflecting the elevation and azimuth angle resolutions, respectively.
\item Next, we study the uplink multi-user mmWave single-sided FD lens MIMO system with limited RF chains at the BS and one single RF chain at each MS. By exploiting the unique array response of the FD lens array at the BS, together with the multi-path sparsity of mmWave channels,  we propose a low-complexity {\it path division multiple access} (PDMA) scheme that is applicable for the general frequency-selective channels. With PDMA, the MSs essentially communicate with the BS via different channel paths without incurring any ISI or inter-user interference (IUI), as long as the AoAs of all MS multi-paths are sufficiently separable at the BS. To this end, a new technique called {\it path delay compensation} is proposed to effectively transform the multi-user frequency-selective MIMO channels to parallel frequency-flat small-size MIMO channels for different MSs, for each of which the low-complexity SC transmission can be applied. For general scenarios where the AoAs cannot be fully separated at the BS, analog beamforming at the MSs and digital combining at the BS are jointly designed to maximize the sum-rate of all MSs  with the MRC (maximal ratio combining) and MMSE (minimum mean square error) based combining, respectively.
\item Last, we propose a new and efficient channel estimation scheme tailored for  PDMA, which consists of three phases, namely {\it power-based antenna selection, path estimation and association, and reduced effective MIMO channel estimation}. Thanks to the angle-dependent energy focusing of lens array at the BS and the low-complexity of PDMA, the proposed channel estimation scheme incurs very small training overhead  in practical mmWave systems and leads to comparable performance as that based on perfect channel state information (CSI). Numerical results are provided to show the effectiveness of the proposed designs, and their sum-rate performance compared to state-of-the-art benchmark systems.
\end{itemize}

 The rest of this paper is organized as follows. Section~\ref{sec:SystemModel} introduces the general FD lens antenna arrays as well as the system model of multi-user single-sided lens MIMO.  In Section~\ref{sec:PDMA}, we propose the PDMA scheme as well as the MRC- and MMSE-based beamforming designs. In Section~\ref{sec:channelEstimation}, an efficient channel estimation scheme tailored for the proposed PDMA scheme is presented. Numerical results are given in Section~\ref{sec:simulation}. Finally, we conclude the paper in Section~\ref{sec:Conclu}.

\emph{Notations:} In this paper, scalars are denoted by italic letters. Boldface lower- and upper-case letters denote vectors and matrices, respectively. $\mathbb{C}^{M\times N}$ denotes the space of $M\times N$ complex-valued matrices, and $\mathbf{I} $ represents an identity matrix.
For an arbitrary-size matrix $\mathbf{A}$,  its complex conjugate, transpose, and Hermitian transpose are denoted by $\mathbf A^*$, $\mathbf{A}^{T}$, and $\mathbf{A}^{H}$, respectively. 
  For a vector $\mathbf a$, $\|\mathbf a\|$ denotes its Euclidean norm. For a non-singular square matrix $\mathbf S$, its matrix inverse is denoted as $\mathbf S^{-1}$. The symbol $j$ represents the imaginary unit of complex numbers, with $j^2=-1$. The notation $\ast$ denotes the linear convolution operation. $\delta(\cdot)$ denotes the Dirac delta function, and $\sinc(\cdot)$ is the ``sinc'' function defined as $\sinc(x)\triangleq \sin(\pi x)/(\pi x)$. For a real number $a$, $\lfloor a \rfloor$ and $\lceil a \rceil$  denote the floor and ceiling operations, respectively. Furthermore, $\mathcal{CN}(\boldsymbol \mu, \mathbf C)$   denotes the circularly symmetric complex-valued Gaussian (CSCG) distributions with mean $\boldsymbol \mu$ and covariance matrix $\mathbf C$. For a set $\mathcal S$, $|\mathcal S|$ denotes its cardinality. Furthermore, $\mathcal S_1 \cap \mathcal S_2$, $\mathcal S_1 \cup \mathcal S_2$ and $\mathcal S_1 \setminus \mathcal S_2$ denote the intersection, union, and complement of two sets $S_1$ and $S_2$, respectively.

\begin{figure}
\centering
\includegraphics[scale=0.5]{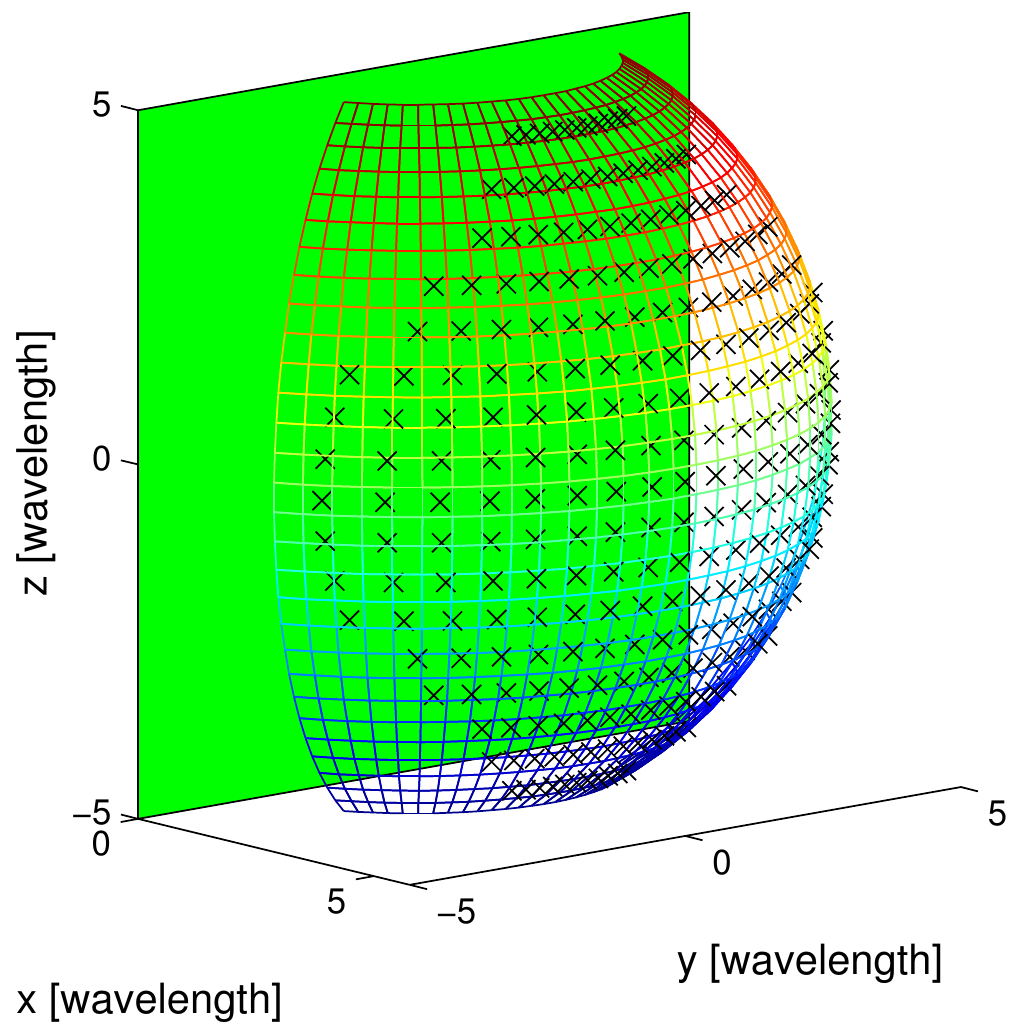}
\caption{An illustration of the FD lens antenna array with $\D_z=\D_y=10$ and azimuth and elevation coverage angles $\Theta=\Phi=120^\circ$. 
\vspace{-3ex}}\label{F:FDLensArrayJSAC}
\end{figure}

\section{System Model}\label{sec:SystemModel}
\subsection{Full-Dimensional Lens Antenna Array}
 Fig.~\ref{F:FDLensArrayJSAC} gives a schematic illustration of a lens antenna array in three-dimensional (3D) coordinate system, which consists of a planar EM lens with negligible thickness in the y-z plane centered at the origin, together with the antenna elements located in the {\it focal surface} of the EM lens. Let $\D_y$ and  $\D_z$ denote the electric dimensions of the lens (i.e., the physical dimensions normalized by signal wavelength) along the y- and z-axis, respectively. The total power captured by the EM lens is then proportional to its effective aperture $\D_y\D_z$. Furthermore, the focal surface of the lens is given by a hemisphere surface around the lens' center with a certain radius $F$, where $F$ is known as the focal length. Therefore, the antenna position of the $m$th array element relative to the lens center can be expressed as $B_m(F\cos \theta_m \cos \phi_m, F\cos \theta_m \sin \phi_m, F\sin \theta_m)$, where $\theta_m\in [-\Theta/2, \Theta/2]$ and $\phi_m \in [-\Phi/2, \Phi/2]$ respectively denote the elevation and azimuth angles corresponding to the location of antenna $m$,  with $\Theta$ and $\Phi$ being the maximum elevation and azimuth  angles to be covered by the lens antenna array. Furthermore, we assume that the antennas are placed  such that $(\theta_m, \phi_m)$ are given by
 \begin{align}
 &\sin\theta_m=\frac{m_e}{\D_z}, \ m_e=0,\pm 1, \cdots, \pm \left\lfloor \D_z \sin \left(\frac{\Theta}{2}\right)\right\rfloor, \label{eq:me} \\
 &\sin \phi_m = \frac{m_a}{\D_y \cos\theta_m},  \ m_a=0, \cdots, \pm \left\lfloor \D_y \cos \theta_m \sin\left(\frac{\Phi}{2}\right)\right\rfloor, \label{eq:ma}
 \end{align}
 where $m_e$ and $m_a$ are referred to as the elevation and azimuth indices of antenna $m$, respectively. Therefore, each antenna $m$ of the lens antenna array is parameterized by both its elevation and azimuth indices as $(m_e, m_a)$.

 For the extreme case of full angle coverage, i.e., $\Theta=\Phi=180^\circ$, the total number of antenna elements $M$ for the lens array with dimension $\D_y \times \D_z$ can be obtained as
 \begin{align}
 M
 &=\sum_{m_e=-\lfloor\D_z\rfloor}^{\lfloor\D_z\rfloor} \left(2\lfloor \D_y \cos\theta_m\rfloor+ 1\right)\notag \\
 & \leq (2\D_y+1)(2\D_z+1)\approx 4\D_y\D_z,
 \end{align}
 where the approximation is made for $\D_y, \D_z\gg 1$. On the other hand, for the conventional UPA of the same dimension $\D_y\times \D_z$ with adjacent elements separated by half wavelength, the required total number of antenna elements can be obtained as $4\D_y\D_z$. Thus, for the same effective array aperture, the FD lens antenna array in general requires less antenna elements as compared to the conventional UPA, thanks to the energy focusing provided  by the EM lens.

 In another extreme case when the lens antenna array is designed by ignoring the signals' elevation angles, we have $\Theta=0$. In this case, all antenna elements are located in the {\it focal arc} as specified by \eqref{eq:ma} with $\theta_m=0$, and the total number of antenna elements with $\Phi=180^\circ$ is $M=2\lfloor\D_y\rfloor+1$. Such a 2D lens array configuration has been studied in our prior work \cite{823}. In the following lemma, we present the array response of the proposed FD lens  antenna array.

\begin{figure}
\centering
\includegraphics[scale=0.38]{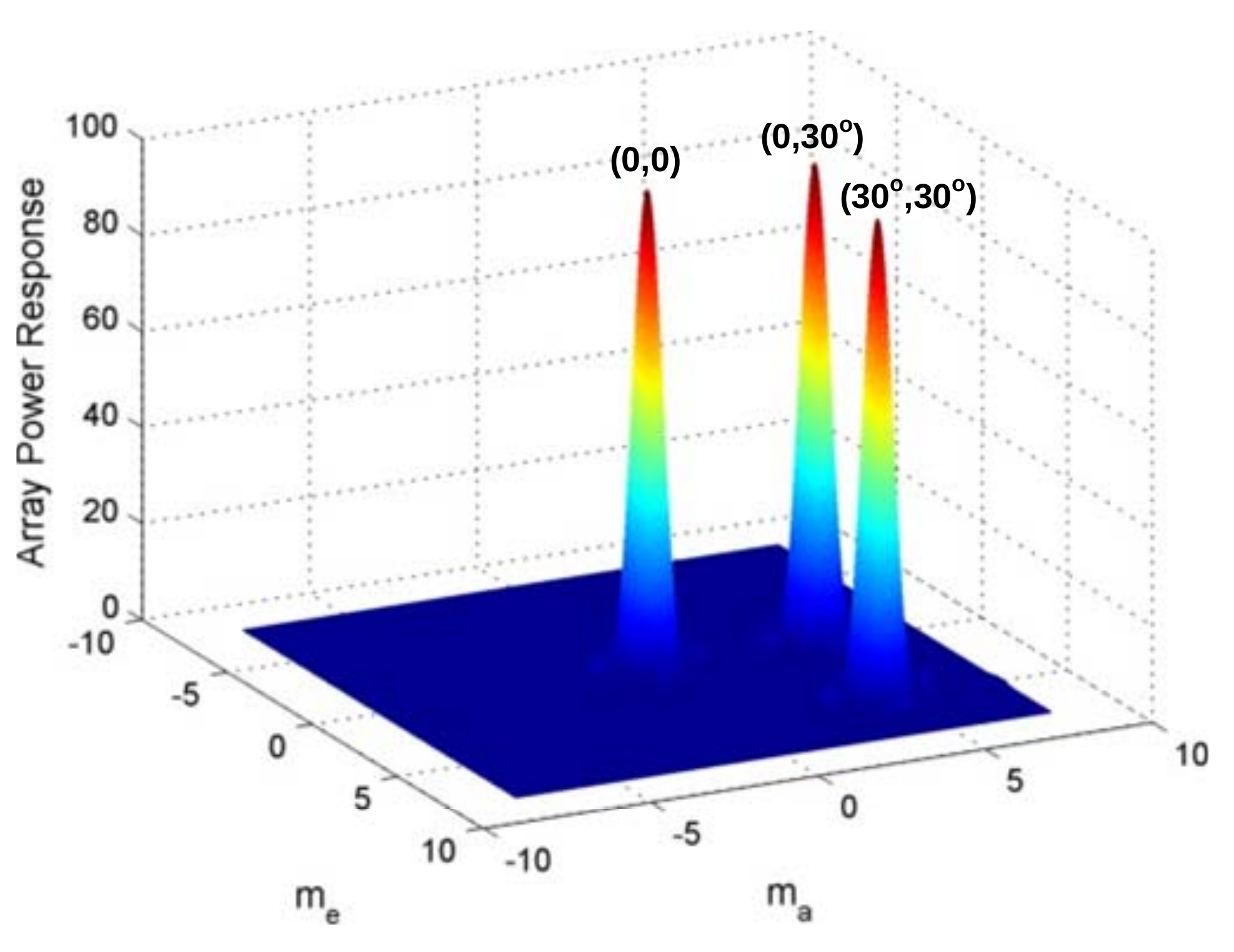}
\caption{Power responses of the FD lens antenna array with three different signal directions.\vspace{-3ex}}\label{F:ArrayResponseJSACLabeled}
\end{figure}

\begin{lemma}\label{lemma:response}
The array response of the FD lens antenna array as a function of the signal's elevation and azimuth angles $(\theta, \phi)$ is expressed as
\begin{align}
a_m(\theta, \phi) = &\sqrt{\D_y \D_z} e^{-j\Phi_0}\sinc (m_e - \D_z \sin \theta)\notag \\
& \times \sinc (m_a - \D_y \cos \theta \sin \phi),  \label{eq:responseFDLens}
\end{align}
where $\Phi_0$ is a common phase shift from the lens' aperture to the array. 
\end{lemma}
\begin{IEEEproof}
Please refer to Appendix~\ref{A:proof}.
\end{IEEEproof}

The array response in \eqref{eq:responseFDLens} is given by a product of two ``sinc'' functions that are related to the elevation and azimuth antenna/angle, respectively. Therefore, for any incident/departure signal from/to a particular direction $(\theta, \phi)$, only those antennas located in the close vicinity of the focal point could receive/steer significant power; whereas the power of all other antennas located far away is almost negligible. As a result, any two simultaneously received/transmitted signals with sufficiently separated directions can be effectively discriminated over different antenna elements. This is illustrated by Fig.~\ref{F:ArrayResponseJSACLabeled} which shows the power responses of the lens antenna array for three different signal directions $(\theta, \phi)=(0,0), (0, 30^\circ), (30^\circ, 30^\circ)$. 
It is observed that the lens antenna array is able to separate signals both in elevation and azimuth directions, where their resolutions can be enhanced by increasing $\D_z$ and $\D_y$, respectively, as can be inferred from \eqref{eq:responseFDLens}. This general lens antenna array design is thus termed {\it FD lens array}. In the special case when the lens antenna array is designed by ignoring the signals' elevation angles, the array response in \eqref{eq:responseFDLens} reduces to that studied in \cite{823}.

A prototype FD lens antenna array has been fabricated, based on which  the preliminary measurement results have verified the angle-dependent energy focusing capabilities of FD lens arrays. The results are reported in \cite{851}.

%

\begin{figure*}
\centering
\includegraphics[scale=0.45]{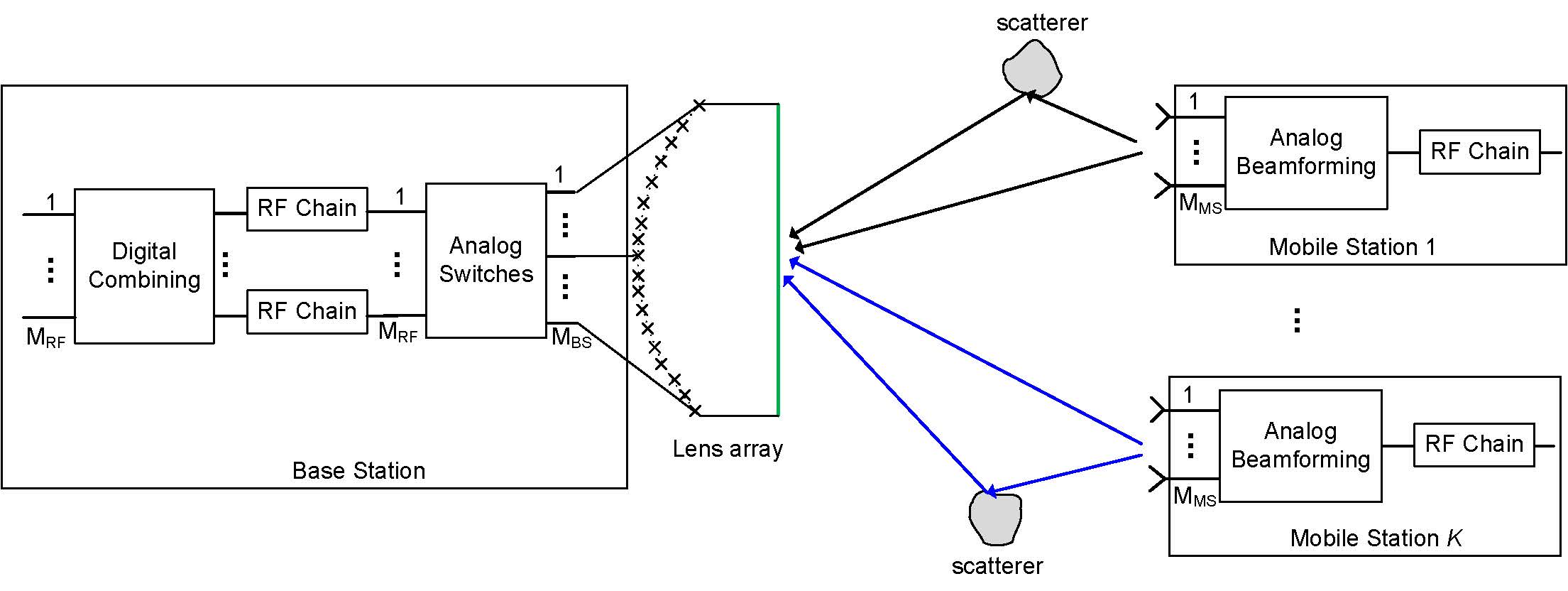}
\caption{Multi-user mmWave MIMO uplink communication with lens array at the BS and conventional array at the MSs.\vspace{-3ex}}\label{F:MultiUserLensMIMO2}
\end{figure*}

\subsection{Multi-User Millimeter Wave  Lens MIMO}
As shown in Fig.~\ref{F:MultiUserLensMIMO2}, we consider a single-cell uplink mmWave MIMO system with $K$ MSs served by a BS. The BS is equipped with an FD lens antenna array with $M_\BS$ antenna elements, and each of the MSs has a conventional antenna array such as UPA with $M_\MS$ elements. For cost-effective implementations, we assume that the BS has only $M_{\RF}$ RF chains, with
$K\leq M_{\RF}< M_\BS$, which are connected to the $M_\BS$ antennas with analog switches. Besides, each of the MSs is equipped with one RF chain only, which is connected to the $M_{\MS}$ antennas with $M_{\MS}$ analog phase shifters. Correspondingly, we assume that {\it digital combining with antenna selection} is performed at the BS receiver and {\it analog beamforming} is performed at the MS transmitters. 
 Furthermore, we assume that the codebook-based analog beamforming is applied at the MSs, i.e., the analog beamforming vector $\mathbf v_k\in \mathbb{C}^{M_\MS\times 1}$ of each MS $k$ is chosen from a pre-determined analog beamforming codebook $\mathcal V=\{\mathbf f_1, \cdots, \mathbf f_{N_{\CB}}\}$, with $N_{\CB}$ denoting the codebook size and $\mathbf f_i\in \mathbb{C}^{M_\MS\times 1}$ being the $i$th analog beamformer with unit norm and all elements having identical amplitude. Note that codebook based analog beamforming has been commonly adopted for short-range mmWave systems \cite{571}, \cite{574}. 

 Under the general multi-path environment, the channel impulse response $\mathbf h_{mk}^{H}(t)\in \mathbb{C}^{1 \times M_\MS}$ from MS $k$ to the $m$th antenna of the BS can be modeled as
 \begin{align}
 \mathbf h_{mk}^{H}(t)=\sum_{l=1}^L \mathbf h_{mkl}^{H} \delta(t-\tau_{kl}), \  m\in \mathcal M, \ k=1,\cdots, K, \label{eq:HkUL}
 \end{align}
 where
 $\mathbf h_{mkl}^H = \alpha_{kl}a_m(\theta_{kl}, \phi_{kl})\mathbf b^H(\theta_{kl}', \phi_{kl}')$ 
  is the $M_\MS$-dimensional vector representing the channel coefficients from MS $k$ to BS antenna $m$ via path $l$, $\mathcal M$ with $|\mathcal M|=M_\BS$ denotes the antenna set of the lens array at the BS, $L$ denotes the maximum number of channel paths 
   which is typically small (e.g., no larger than 3) due to the multi-path sparsity of mmWave channels \cite{569}, $\alpha_{kl}$ and $\tau_{kl}$ denote the complex-valued path gain and the delay for the $l$th path, respectively, $\theta_{kl}$ and $\phi_{kl}$ are the elevation and azimuth AoA,  respectively, and $\theta_{kl}'$ and $\phi_{kl}'$ are the elevation and azimuth AoD, respectively. Furthermore, $a_m(\cdot)$ is the lens array response at the BS as given in \eqref{eq:responseFDLens},  and $\mathbf b(\cdot) \in \mathbb{C}^{M_\MS \times 1}$ is the array response of the conventional array at the MS. 


   \section{Path Division Multiple Access}\label{sec:PDMA}
In this section, we assume perfect CSI at the BS as well as at the MSs, while the channel estimation scheme will be studied in Section~\ref{sec:channelEstimation}. By exploiting the unique lens array response in \eqref{eq:responseFDLens}, together with the  {multi-path sparsity} of mmWave channels, we propose a PDMA scheme for multi-user mmWave lens MIMO systems that is applicable for the general wide-band frequency-selective channels. With this scheme, the MSs essentially communicate with the BS via distinct channel paths of different AoAs using the low-complexity SC transmission. As long as the lens array at the BS is sufficiently large to resolve all the multi-path signals from all MSs, PDMA is able to inherently eliminate both the ISI and IUI with only limited number of RF chains and low signal processing complexity. 

With SC transmission, let $s_k[n]$ be the independent information-bearing symbols sent by MS $k$, with $n$ denoting the symbol index. 
 The transmitted signal vector $\mathbf x_k[n]\in \mathbb{C}^{M_{\MS}\times 1}$ by the $M_{\MS}$ antennas of MS $k$ is given by
\begin{align}
\mathbf x_k[n] = \sqrt{p_k} \mathbf v_k s_k[n], \ k=1,\cdots, K,
\end{align}
where $p_k$ denotes the transmit power of MS $k$ and $\mathbf v_k \in \mathcal V$ is its analog beamforming vector chosen from the codebook $\mathcal V$. Note that only one data stream can be transmitted by each MS since it has only one RF chain.

  Since only $M_\RF<M_{\BS}$ RF chains are available at the BS, antenna selection needs to be applied. Denote by $\mathcal M_S\subset \mathcal M$ with $|\mathcal M_S|=M_\RF$ the subset of the selected BS antennas based on certain selection schemes (e.g., the low-complexity power-based antenna selection given in Section~\ref{sec:channelEstimation}). Note that different from the conventional antenna systems, power-based antenna selection works particularly well for mmWave lens array, thanks to its energy focusing property. In particular, if $KLJ\leq M_\RF <M_{\BS}$, with $J$ denoting the maximum number of energy focusing antennas with non-negligible power for each path, then antenna selection at the BS incurs essentially no power loss.
   The signal received by the selected BS antennas can be expressed as
 \begin{align}
 y_m&[n]=\sum_{k=1}^K \mathbf h_{mk}^H[n] \ast \mathbf x_k[n] + z_m[n] \notag \\
 &\hspace{-3ex}= \sum_{k=1}^K \sum_{l=1}^L \mathbf h_{mkl}^H \mathbf v_k \sqrt{p_k} s_k[n-n_{kl}] + z_m[n], m\in \mathcal M_S, \label{eq:ym}
 \end{align}
 where  $\mathbf h_{mk}^H[n]=\sum_{l=1}^L \mathbf h_{mkl}^H\delta[n-n_{kl}]$ denotes the discrete-time equivalent of the channel impulse response in \eqref{eq:HkUL}, with $n_{kl}$ denoting the path delay in symbol durations, and $z_m[n]\sim \mathcal{CN}(0, \sigma^2)$ is the independent CSCG noise at the BS antenna $m$ with zero mean and power $\sigma^2$. Note that we assume the general wide-band channels such that $\forall k$, $n_{kl}\neq n_{kl'}$, $\forall l\neq l'$. In the special case when certain paths of each user have identical delays (but different AoA/AoD without loss of generality), the following schemes can be directly applied by grouping those paths with identical delays.

It is observed from \eqref{eq:ym} that the received signal $y_m[n]$ by each BS antenna $m$ is a superposition of the $K$ data streams, each arriving via $L$ multi-paths with different delays. In other words, \eqref{eq:ym} essentially constitutes a {\it weighted} linear combination of $KL$ independent data symbols, where the weight of the $(k,l)$th symbol is given by $\mathbf h_{mkl}^H\mathbf v_k \sqrt{p_k}$.  As a consequence, the desired signal in $y_m[n]$ for each MS in general suffers from both the ISI and IUI. However, a close look at the expression for $\mathbf h^H_{mkl}$ reveals that, out of all the $KL$ multi-paths, only those having the AoA $(\theta_{kl}, \phi_{kl})$ with the energy focusing antenna around $m$ would have non-negligible signal contribution in $y_m[n]$, thanks to the ``sinc'' array response given in \eqref{eq:responseFDLens}. This thus offers a unique opportunity for intrinsic ISI and IUI suppression, by equipping the BS with a sufficiently large lens antenna array with sufficiently fine angle resolution to resolve all the $KL$ multi-paths. As a result, each MS may virtually communicate with the BS orthogonally via the non-overlapping energy focusing antennas corresponding to the different paths, thus termed PDMA.

For the generic systems where the BS is unable to perfectly resolve all the $KL$ paths, we propose two PDMA receiver combining techniques for \eqref{eq:ym} based on MRC and MMSE, respectively. For ease of presentation, let $\mathcal L \triangleq \{(k,l): k=1,\cdots, K, l=1,\cdots, L\}$ be the set of all the $KL$ paths. Furthermore, define
\begin{align}
\beta_{mkl}\triangleq \alpha_{kl}a_m(\theta_{kl}, \phi_{kl}), \forall m\in \mathcal M_S, \ (k, l)\in \mathcal L. \label{eq:betamkl}
\end{align}
Physically, $|\beta_{mkl}|^2$ signifies the relative signal power received at BS antenna $m\in \mathcal M_S$ from path $(k,l)\in \mathcal L$ if all MSs transmit omni-directionally with identical power.



\subsection{MRC-based PDMA}
With MRC-based PDMA, each BS antenna is synchronized to its strongest path, and the simple MRC receive combining is applied for detecting the signal from each MS based on the effective MIMO channel after applying a new technique called {\it path delay compensation}. Specifically, for $m\in \mathcal M_S$, denote by $(k_m, l_m)$ the strongest path out of all the $KL$ paths, i.e.,
$(k_m, l_m) =\mathrm{\arg} \underset{(k,l)\in \mathcal L}{\max} |\beta_{mkl}|^2$ and $\bar{\mathcal L}_m \triangleq \mathcal L \setminus (k_m, l_m)$ the set of the remaining $(KL-1)$ paths. Then by synchronizing the receiver of BS antenna $m\in \mathcal M_S$ to path $(k_m, l_m)$, $y_m[n]$ in \eqref{eq:ym} can be decomposed as
\begin{align}
y_m[n] =&\mathbf h_{mk_ml_m}^H \mathbf v_{k_m}\sqrt{p_{k_m}}s_{k_m}[n-n_{k_ml_m}] \notag \\
&+ \underbrace{\sum_{(k,l)\in \bar{\mathcal L}_m} \mathbf h_{mkl}^H \mathbf v_k \sqrt{p_k}s_k[n-n_{kl}]}_{\text{IPI}}+z_m[n], \label{eq:ymn4}
\end{align}
where the second term in \eqref{eq:ymn4} consists of signals from all other $KL-1$ paths at antenna $m$, thus causes the inter-path interference (IPI), including both the ISI and IUI.  With perfect {\it path delay compensation} at each antenna $m\in \mathcal M_S$, i.e., by letting $\bar y_m[n] \triangleq y_m[n+n_{k_ml_m}]$, \eqref{eq:ymn4} is equivalent to
\begin{align}
\bar y_m[n] = \mathbf h_{mk_ml_m}^H \mathbf v_{k_m}\sqrt{p_{k_m}}s_{k_m}[n]+ \bar{z}_{m,\text{IPI}}[n]+ \bar{z}_m[n],  \label{eq:barym}
\end{align}
where $\bar z_{m,\text{IPI}}[n]$ represents the IPI term in \eqref{eq:ymn4}, and $\bar{z}_m[n]\triangleq z_m[n+n_{k_ml_m}]$ has the identical distribution as $z_m[n]$.

For each MS $k$, let $\mathcal M_k\subset \mathcal M_S$ be the subset of the selected BS antennas that receive the strongest multi-path power from MS $k$, i.e., $\mathcal M_k \triangleq \{m\in \mathcal M_S: k_m=k\}$. Note that if $M_{\RF}$ is too small, $\mathcal M_k$ could be empty for those MSs that are far away from the BS and thus have weak channels. This is desirable from a sum-rate maximization perspective, but may cause the user fairness issue since the distant MSs may have very low or even zero data rate. Such an issue could be mitigated via proper user scheduling and/or power control, which is out of the scope of this paper. Since each antenna $m$ is associated with only one of the $KL$ paths, we thus have $\mathcal M_k \bigcap \mathcal M_{k'}=\emptyset$, $\forall k\neq k'$.  For each MS $k$, by concatenating $\bar y_m[n]$, $\forall m\in \mathcal M_k$, as $\mathbf r_k[n]\in \mathbb{C}^{|\mathcal M_k|\times 1}$, \eqref{eq:barym} can be equivalently written as
\begin{align}
 \hspace{-2ex} \mathbf r_k[n] =  \mathbf G_k \mathbf v_k \sqrt{p_k}s_k[n] + \mathbf z_{k,\text{IPI}}[n]+ \mathbf z_k[n], \ \forall k, \label{eq:rk}
\end{align}
where $\mathbf G_k \in \mathbb{C}^{|\mathcal M_k|\times M_\MS}$ is the effective frequency-flat MIMO channel matrix after path delay compensation for MS $k$ with rows $\mathbf h_{mkl_m}^H$, $m\in \mathcal M_k$, $\mathbf z_{k,\text{IPI}}[n]$ and $\mathbf z_k[n]\in \mathbb{C}^{|\mathcal M_k|\times 1}$ denote the IPI and noise vector, respectively.

Denote by $\mathbf u_k  \in \mathbb{C}^{|\mathcal M_k|\times 1}$ with $\|\mathbf u_k\|=1$ the receive combining vector that is applied to \eqref{eq:rk} for detecting $s_k[n]$. With the MRC-based scheme, the IPI in \eqref{eq:rk} is not taken into account for the combiner design, and the transmit beamforming and receive combining vectors $\{\mathbf v_k, \mathbf u_k\}_{k=1}^K$ are designed to simply maximize the desired signal power for each MS $k$, i.e.,
\begin{align}
(\mathbf v_k^\star, \mathbf u_{k}^\star) = \mathrm{arg} \underset{\|\mathbf u_{k}\|=1, \mathbf v_k \in \mathcal V}{\max} & |\mathbf u_{k}^H \mathbf G_k \mathbf v_k|^2. \label{eq:BF1}
\end{align}
It can be shown that the optimal solution to \eqref{eq:BF1} is
\begin{align}
  \mathbf v_k^\star=\mathrm{arg} \underset{\mathbf v_k \in \mathcal V}{\ \max} \|\mathbf G_{k}  \mathbf v_k\|^2, \ \mathbf u_{_k}^\star=\frac{\mathbf G_{k} \mathbf v_k^\star}{\|\mathbf G_{k}  \mathbf v_k^\star\|}, \label{eq:uk1vk1}
\end{align}
which can be obtained by comparing all the $N_{\CB}$ analog beamforming vectors in $\mathcal V$.

While the MRC-based design in \eqref{eq:uk1vk1} is sub-optimal in general due to the ignoring of the IPI term in \eqref{eq:rk}, it is asymptotically optimal when $M_{\BS}\gg KL$  and $M_\RF\geq KL$ since in this case, the IPI term vanishes. Specifically, with $M_\BS \gg KL$,  it is of high probability that each antenna $m\in \mathcal M_S$ receives non-negligible signal power from at most one path. As a result, $\forall (k,l)\in \bar {\mathcal L}_m$, we have $\beta_{mkl}\approx 0$ and hence $\mathbf h_{mkl}\approx \mathbf 0$.  Thus, the IPI term in \eqref{eq:ymn4} vanishes, and \eqref{eq:rk} reduces to  $K$ parallel frequency-flat small MIMO channels without any interference. Hence, maximizing the desired signal power as in \eqref{eq:uk1vk1} is optimal in this case.

\begin{figure*}
\centering
\includegraphics[scale=0.55]{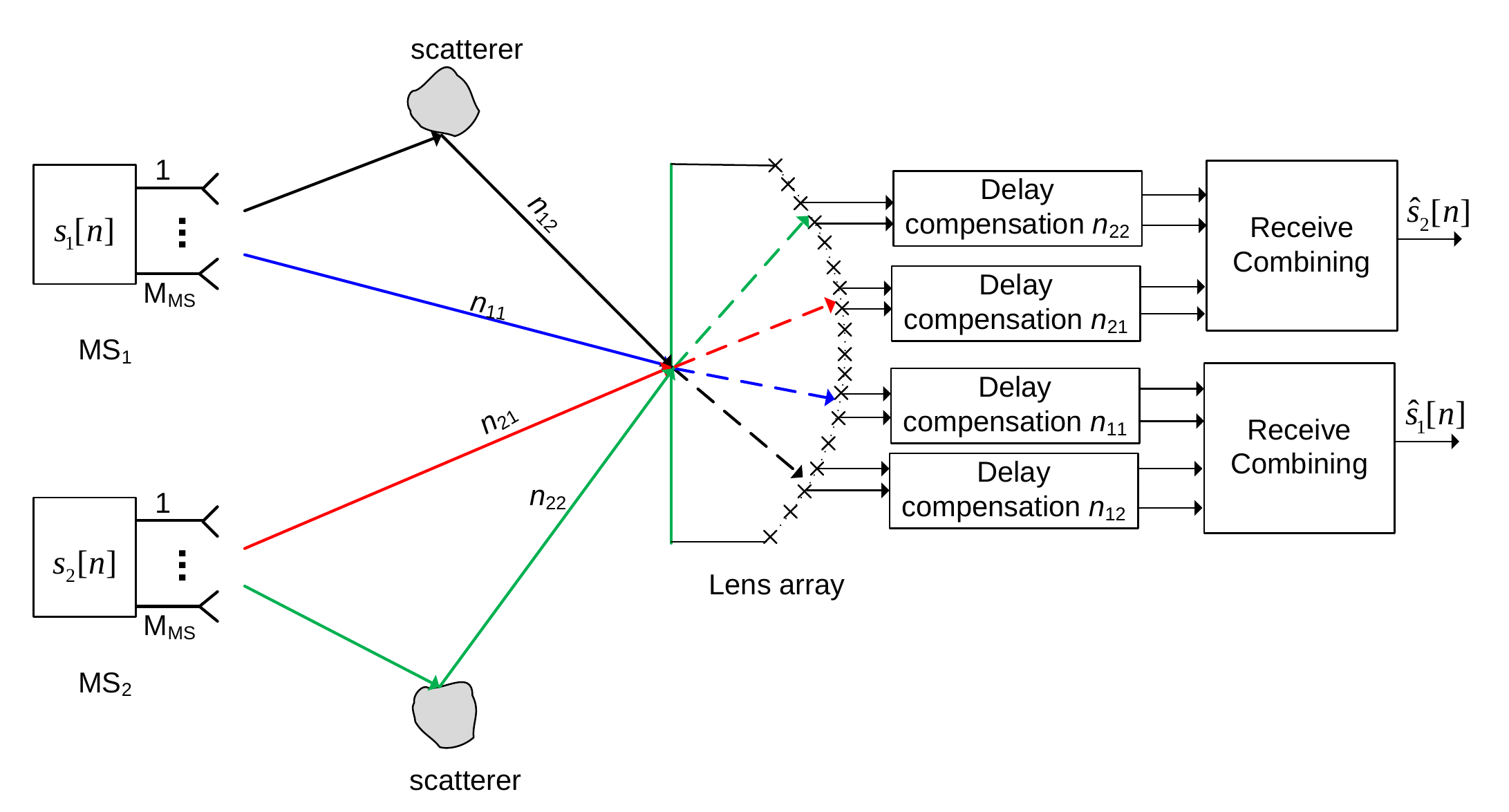}
\caption{Illustration of MRC-based PDMA with sufficiently separated AoAs, where $n_{kl}$ represents the path delay for the $l$th path of MS $k$.}\label{F:UplinkPDMA}
\end{figure*}

The proposed  MRC-based PDMA scheme with sufficiently separated AoAs is illustrated in Fig.~\ref{F:UplinkPDMA} for the case of $K=2$, $L=2$ and $M_\RF=8$.

\subsection{MMSE-based PDMA}
In this subsection, we propose the MMSE-based PDMA scheme with the ISI and IUI further mitigated.
Under insufficient AoA separations, each antenna $m\in \mathcal M_S$ may potentially receive desired signal (as well as interference) for up to $K$ MSs. Thus, for detecting the data stream $s_k[n]$ of each MS $k$,  all antennas $m\in \mathcal M_S$ should be utilized. Specifically, for detecting $s_k[n]$,  antenna $m\in \mathcal M_S$ is synchronized to the strongest path, denoted as $l_m^k\in \{1,\cdots, L\}$, among all the $L$ multi-paths of MS $k$, where
$l_m^k =  \mathrm{arg}\underset{l=1,\cdots, L}{\max} \ |\beta_{mkl}|^2, \ m\in \mathcal M_S$.
With the received signal $y_m[n]$ in \eqref{eq:ym} compensated by the corresponding delay $n_{kl_m^k}$ by letting $y_{mk}[n]\triangleq y_m[n+n_{kl_m^k}]$, the resultant signal for MS $k$ at antenna $m$ can be written as
\begin{align}
y_{mk}&[n] = \underbrace{\mathbf h_{mkl_m^k}^H \mathbf v_k \sqrt{p_k} s_k[n]}_{\text{desired signal}} + \underbrace{\sum_{l\neq l_{m}^k}^L \mathbf h_{mkl}^H \mathbf v_k \sqrt{p_k} s_k[n- \Delta_{kl, kl_m^k}]}_{\text{ISI}} \notag \\
& \hspace{-3ex} +\underbrace{\sum_{k'\neq k}^K \sum_{l=1}^L \mathbf h_{mk'l}^H \mathbf v_{k'} \sqrt{p_{k'}} s_{k'}[n-\Delta_{k'l, kl_m^k}]}_{\text{IUI}} + z_{m}[n+n_{kl_m^k}], \label{eq:ymk}
\end{align}
where $\Delta_{k'l',kl}\triangleq n_{k'l'}-n_{kl}$ denotes the {\it excessive path delay} between the $l$th path of MS $k$ and the $l'$th path of MS $k'$. Let $\mu$ denote the maximum path delay in symbol durations for all the $KL$ paths, i.e., $0\leq n_{kl}\leq \mu$, $\forall k, l$. We then have $\Delta_{k'l',kl}\in \{0, \pm 1, \cdots, \pm \mu\}$. To derive the signal-to-interference-plus-noise ratio (SINR) expression for MS $k$, the signals in \eqref{eq:ymk} need to be reformulated by grouping those interfering symbols with identical excessive delays since they are correlated if originated from the same MS. To this end, for excessive delay $-\mu \leq i \leq \mu$, $m\in \mathcal M_S$, and $k, k'=1,\cdots, K$, we define
\begin{align} \label{eq:gmkk}
\mathbf g_{m, kk'}^H[i]\triangleq
\begin{cases}
\mathbf h_{mk'l}^H, \ & \text{if } \exists l\in\{1,\cdots, L\} \text{ s.t. } n_{k'l}-n_{kl_m^k}=i, \\
\mathbf 0, \ & \text{otherwise.}
\end{cases}
\end{align}
Then \eqref{eq:ymk} can be equivalently written as
\begin{align}
\hspace{-2ex} y_{mk}[n]&=  \mathbf g_{m,kk}^H[0]\mathbf v_k \sqrt{p_k}s_k[n] + \sum_{i=-\mu, i\neq 0}^\mu \mathbf g_{m,kk}^H[i] \mathbf v_k \sqrt{p_k}s_k[n-i]\notag  \\
& + \sum_{k'\neq k}^K \sum_{i=-\mu}^{\mu} \mathbf g_{m,kk'}^H[i]\mathbf v_{k'}\sqrt{p_{k'}}s_{k'}[n-i]+ z_m[n], \ m\in \mathcal M_S. \label{eq:ymk2}
\end{align}
By concatenating the signals of all antennas in $\mathcal M_S$, \eqref{eq:ymk2} can be compactly written as
\begin{align}
\bar {\mathbf y}_k[n] = &\mathbf G_{kk}[0]\mathbf v_k \sqrt{p_k}s_k[n] + \sum_{i=-\mu, i\neq 0}^\mu \mathbf G_{kk}[i] \mathbf v_k \sqrt{p_k}s_k[n-i] \notag \\
& + \sum_{k'\neq k}^K \sum_{i=-\mu}^\mu \mathbf G_{kk'}[i]\mathbf v_{k'}\sqrt{p_{k'}}s_{k'}[n-i]+\mathbf z[n],
\end{align}
where $\bar {\mathbf y}_k[n]$ is an $|\mathcal{M}_S|\times 1$ vector with elements given by $y_{mk}[n]$, $m\in \mathcal M_S$, and $\mathbf G_{kk'}[i]\in \mathbb{C}^{|\mathcal M_S|\times M_\MS}$ with the rows given by $\mathbf g_{m,kk'}[i]$, $m\in \mathcal M_S$, is the effective MIMO channel matrix from MS $k'$ to MS $k$ with excessive path delay $i$. For detecting $s_k[n]$ for MS $k$, a receive combining vector $\mathbf {\bar u}_k\in \mathbb{C}^{|\mathcal M_S|\times 1}$ with $\|\bar{\mathbf u}_k\|=1$ is applied to $\bar {\mathbf y}_k[n]$, which yields
\begin{align}
\hat{s}_k&[n]  = \bar{\mathbf u}_k^H\bar{\mathbf y}_k[n]=  \underbrace{\bar{\mathbf u}_k^H \mathbf G_{kk}[0]\mathbf v_k \sqrt{p_k}s_k[n]}_{\text{desired signal}} \notag \\
 &+ \underbrace{\sum_{i=-\mu, i\neq 0}^\mu \bar{\mathbf u}_k^H \mathbf G_{kk}[i] \mathbf v_k \sqrt{p_k}s_k[n-i]}_{\text{ISI}} \notag \\
& + \underbrace{\sum_{k'\neq k}^K \sum_{i=-\mu}^\mu \bar{\mathbf u}_k^H \mathbf G_{kk'}[i]\mathbf v_{k'}\sqrt{p_{k'}}s_{k'}[n-i]}_{\text{IUI}}+\bar{\mathbf u}_k^H \mathbf z[n].\label{eq:hatsk}
\end{align}

By treating the ISI and the IUI as noise, the resultant SINR of MS $k$ can be expressed as \eqref{eq:gammak} shown at the top of the next page,
\begin{figure*}
\begin{align}
\gamma_k=\frac{p_k|\bar{\mathbf u}_k^H \mathbf G_{kk}[0]\mathbf v_k|^2}{\sum_{i=-\mu, i\neq 0}^\mu p_k |\bar{\mathbf u}_k^H \mathbf G_{kk}[i]\mathbf v_k|^2+\sum_{k'\neq k}^K \sum_{i=-\mu}^\mu p_{k'}|\bar{\mathbf u}_k^H \mathbf G_{kk'}[i]\mathbf v_{k'}|^2+\sigma^2},\label{eq:gammak}
\end{align}
\end{figure*}
where we have used the fact that the data symbols $s_k[n]$ and $s_{k'}[n']$ are independent for $k\neq k'$ or $n\neq n'$.
As a result, an achievable data rate for MS $k$ is $R_k=\log_2(1+\gamma_k)$ assuming Gaussian input, $\forall k$. The transmit and receive beamforming vectors $\{\mathbf v_k, \bar{\mathbf u}_k\}_{k=1}^K$ can be optimized for sum-rate maximization by solving the following optimization problem
\begin{align}
\mathrm{(P1)} \quad \underset{\mathbf v_k\in \mathcal V, \|\bar{\mathbf u}_k\|=1, \forall k}{\max} \ & \sum_{k=1}^K \log_2\left( 1+ \gamma_k \right). 
\end{align}
(P1) is a discrete optimization problem, which incurs an exponential complexity $O(N_{\CB}^K)$ if the exhaustive search method is applied for finding the optimal solution.

We thus propose an efficient suboptimal solution to (P1) by firstly obtaining the best transmit beamforming vector $\mathbf v_k$ by ignoring the ISI and IUI terms in \eqref{eq:gammak}, and then designing the receive combining vector $\bar{\mathbf u}_k$ with the interference properly mitigated. Specifically, by ignoring the ISI and IUI,  $\mathbf v_k$ is designed to maximize the numerator of \eqref{eq:gammak} as
\begin{align}
\mathbf v_k^\star= \mathrm{arg} \underset{\mathbf v_k \in \mathcal V}{\ \max} \|\mathbf G_{kk}[0]  \mathbf v_k\|^2, \forall k, \label{eq:vk2}
\end{align}
which can be obtained by comparing all analog beamformers in $\mathcal V$ with complexity $O(N_{\CB})$. With the transmit beamforming vectors fixed as in \eqref{eq:vk2}, the optimal receive combining vector to (P1) is given by the MMSE solution, i.e,
\begin{align}
\bar{\mathbf u}_k^\star=\frac{\mathbf C_k^{-1} \mathbf G_{kk}[0]\mathbf v_{k}^\star }{\big \|\mathbf C_k^{-1}\mathbf G_{kk}[0]\mathbf v_{k}^\star \big \|}, \forall k, \label{eq:uk2}
\end{align}
where
\begin{align}
\mathbf C_k=&\sum_{i=-\mu, i\neq 0}^\mu p_k \mathbf G_{kk}[i]\mathbf v_k^\star \mathbf v_k^{\star H} \mathbf G_{kk}^H[i]\notag \\
&+\sum_{k'\neq k}^K \sum_{i=-\mu}^\mu p_{k'}\mathbf G_{kk'}[i]\mathbf v_{k'}^\star \mathbf v_{k'}^{\star H} \mathbf G_{kk'}^H[i]+\sigma^2 \mathbf I
\end{align}
denotes the covariance matrix of the interference-plus-noise term for MS $k$. The resultant SINR for MS $k$ is
\begin{align}
\gamma_k = p_k \mathbf v_k^{\star H} \mathbf G_{kk}^H[0] \mathbf C_k^{-1} \mathbf G_{kk}[0]\mathbf v_k^\star.
\end{align}


If all the $KL$ multi-paths are well separated at the BS, it then follows from \eqref{eq:gmkk} that $\mathbf G_{kk'}[i]\approx \mathbf 0$, $\forall k'\neq k$ or $i\neq 0$. As a result, both the ISI and IUI in \eqref{eq:hatsk} disappear and it can be shown that the beamforming solution in \eqref{eq:vk2} and \eqref{eq:uk2} is equivalent to the MRC-based solution in \eqref{eq:uk1vk1}. 

\section{Channel Estimation}\label{sec:channelEstimation}
 In this section, we propose a novel channel estimation scheme tailored for the MRC-based PDMA scheme in single-sided lens MIMO system.\footnote{The channel estimation scheme for the MMSE-based PDMA follows the same protocol as that for MRC-based PDMA, but is more complicated as more channel coefficients need to be estimated, which is thus omitted due to the space limitations.}  Note that for obtaining the complete knowledge of the channel impulse response of each MS $k$,  the BS in general needs to estimate the corresponding  $M_{\BS}\times M_{\MS}$ MIMO channel matrices for each of the $\mu+1$ taps, where $\mu$ denotes the maximum path delay in symbol durations. Furthermore, with only $M_{\RF}< M_{\BS}$ RF chains at the BS, it can be shown that such a brute-force channel estimation scheme requires a minimum training duration $T'=\lceil\frac{M_{\BS}}{M_{\RF}}\rceil \mu KM_\MS$, which is prohibitive in practice. For example, for a typical system with $M_{\BS}=317$ (corresponding to $\Theta=\Phi=180^\circ$ and $\D_y=\D_z=10$), $M_{\RF}=10$, $\mu=50$, $K=5$, and $M_\MS=16$, we have $T'=128000$. On the other hand, if the channel coherence time is $0.1$ms, and the mmWave bandwidth is $500$MHz,  the total number of symbols within the channel coherence time is approximately $T_c=50000$. Thus, the required training duration with the brute-force training is even much longer than the channel coherence time, which is obviously impractical. In this section, by exploiting the angle and path sparsity of the mmWave channels jointly with the lens array enabled MRC-based PDMA scheme proposed in the preceding section,  we propose an efficient channel estimation scheme with much reduced training overhead, which consists of three phases, namely {\it power-based antenna selection, path estimation and association, and reduced MIMO channel estimation}.



%
\subsection{Power Based Antenna Selection at the BS}
Since only $M_\RF<M_{\BS}$ RF chains are available at the BS, it needs to select $M_\RF$ antennas at the beginning of each channel coherence block. Thanks to the AoA-dependent energy focusing of lens antenna array as well as the limited number of multi-path for mmWave channels, it is intuitive that only those antennas located around the energy focusing regions of the multi-path signals should be selected, which can be attained based on the low-complexity power probing by the MSs. Specifically, in the first phase of channel estimation, the $K$ MSs simultaneously send {\it identical} power-probing symbols {\it omni-directionally} to the BS, i.e., $s_k[n]=s, \forall k=1,\cdots, K, n=1, \cdots, T_1$, and $\mathbf v_k=\bar{\mathbf f}$, $\forall k$, where $T_1$ denotes the training duration of the first phase, $\bar{\mathbf f}$ represents the omni-directional analog beamformer applied by  the MS with $\mathbf b^H(\phi, \theta)\bar{\mathbf f}\approx C$, $\forall \theta, \phi$. By discarding the first $\mu$ symbol durations during which not all multi-path signals have  arrived, the received signal at BS antenna $m\in \mathcal M$ can be written as
\begin{align}
r_m[n]&=\sum_{k=1}^K \sum_{l=1}^L \mathbf h_{mkl}^H \mathbf{\bar f} \sqrt{p_{k}}s_{k}[n-n_{kl}] + z_m[n]\notag \\
&\approx \sqrt{p_\tr} Cs \sum_{k=1}^K \sum_{l=1}^L \alpha_{kl} a_m(\theta_{kl}, \phi_{kl})+z_m[n], \notag \\
 & \hspace{5ex} n=\mu+1, \cdots, T_1, \label{eq:rmn}
\end{align}
where $p_\tr$ denotes the training power by each MS.
Note that since identical power probing symbols are sent, i.e., $s_k[n]=s$, $\forall n$,  the BS essentially receives the same signal across $n$ (without considering the noise). At the BS side, the $M_{\RF}$ RF chains   sequentially scan over the $M_{\BS}$ antennas to obtain their respectively received power levels $Q_m=|r_m[n]|^2$, $m\in \mathcal M$. After that, those $M_{\RF}$ antennas with the highest power levels are selected, which are denoted by the set $\mathcal M_{S}$. Due to the ``sinc''-type array response $a_m(\cdot)$ given in \eqref{eq:responseFDLens}, it follows from \eqref{eq:rmn} that those antennas in the vicinity of the energy focusing regions of the $KL$ multi-paths are more likely to be selected, as desired. Furthermore, since the RF chain scanning process requires duration $\lceil \frac{M_{\BS}}{M_{\RF}}\rceil$, the minimum required training duration for antenna selection in phase 1 can be obtained as $T_1=\lceil \frac{M_{\BS}}{M_{\RF}}\rceil+\mu$.
%
%
%
%
\subsection{Path Estimation and Association at the BS}
\begin{figure*}
\centering
\includegraphics[scale=0.7]{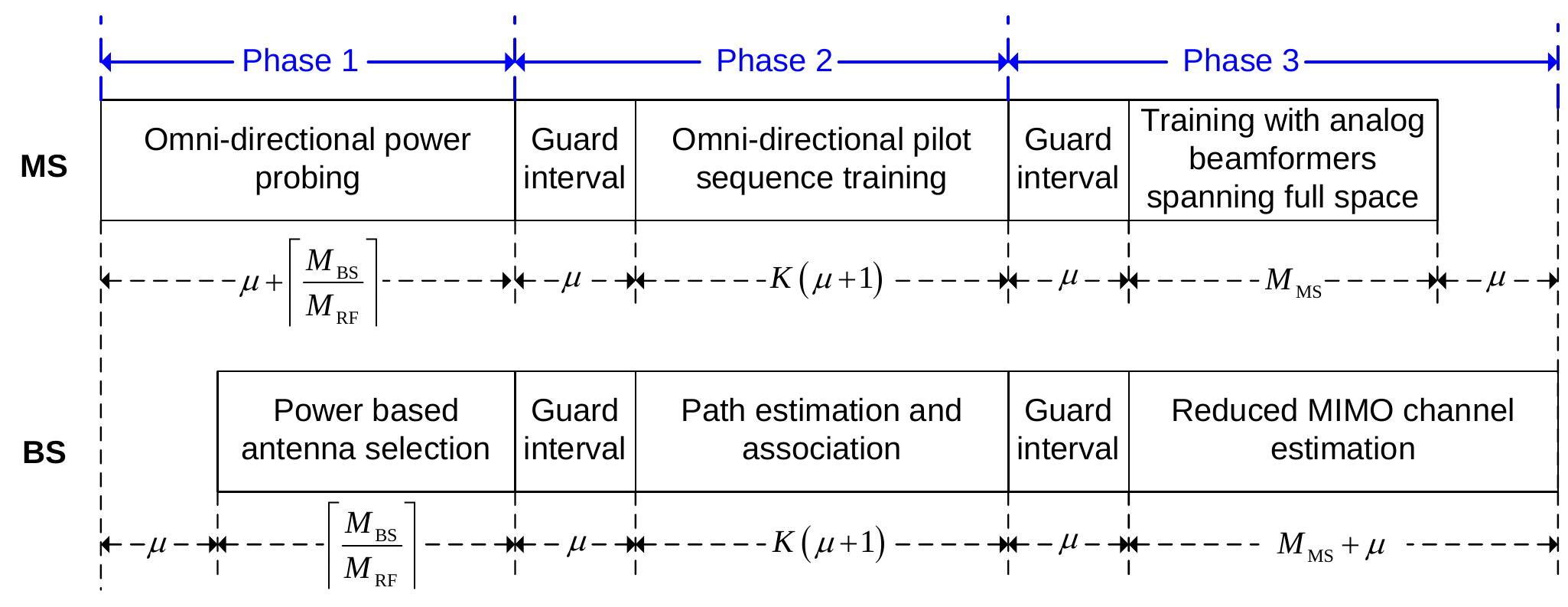}
\caption{Illustration of the proposed channel estimation scheme for mmWave lens MIMO.}\label{F:ChannelEstimationFrameStructure}
\end{figure*}

The next training phase aims to associate each of the selected BS antennas $m\in \mathcal M_S$ with one of the $KL$ paths (if any) for delay compensation and subsequent data reception. To this end, the relative strengths of the received signals via the different paths at antenna $m$ as well as their path delays are estimated. To avoid receiving the power training symbols sent in phase 1, a guard time of interval $\mu$ is inserted at the beginning of phase 2, as shown in Fig.~\ref{F:ChannelEstimationFrameStructure}. After that, each   MS $k$ sends a pilot sequence $s_k[n]$, $n=1,\cdots, T_2$, with the omni-directional analog beamforming vector $\bar{\mathbf f}$, where $T_2$ denotes the training duration of phase 2. The signal received by each of the selected BS antennas $m\in \mathcal M_S$ can be written as
\begin{align}
r_m&[n]\approx \sqrt{p_\tr}C  \sum_{k=1}^K \sum_{l=1}^L \alpha_{kl} a_m(\theta_{kl}, \phi_{kl})s_k[n-n_{kl}] + z_m[n] \notag\\
&= \sqrt{p_\tr}C \sum_{k=1}^K\sum_{l=1}^L \beta_{mkl}s_k[n-n_{kl}]+z_m[n], \ n=1,\cdots, T_2,\label{eq:rmn5}
\end{align}
where $\beta_{mkl}$ is defined in \eqref{eq:betamkl}
signifying the relative path gain of path $(k,l)$ to BS antenna $m$ with omnidirectional transmission by  the MSs. Based on the received signal $r_m[n]$ in \eqref{eq:rmn5}, the BS antenna $m$ needs to estimate the effective path gain $\beta_{mkl}$ as well as the path delay $n_{kl}$. To this end, \eqref{eq:rmn5} is reformulated in terms of all the $\mu+1$ taps. Specifically, since $0\leq n_{kl} \leq \mu$, $\forall k, l$, for tap delay $0\leq i \leq \mu$, define
\begin{align}
\beta_{mk}[i]=
\begin{cases}
\beta_{mkl}, \ & \text{ if } \exists l\in \{1,\cdots, L\} \text { such that } n_{kl}=i, \\
0, \ & \text{ otherwise.}
\end{cases}
\end{align}
As a result, \eqref{eq:rmn5} can be equivalently written as
\begin{align}
r_m[n]&=\sqrt{p_\tr}C\sum_{k=1}^K \sum_{i=0}^\mu \beta_{mk}[i]s_k[n-i]+z_m[n], \notag \\
&=\sqrt{p_\tr}C \sum_{i=0}^\mu \mathbf s^T[n-i] \boldsymbol \beta_{m}[i] +z_m[n], \  n=1,\cdots, T_2.\label{eq:rmn6}
\end{align}
where $\boldsymbol \beta_{m}[i]\triangleq \left[\beta_{m1}[i], \cdots, \beta_{mK}[i] \right]^T\in \mathbb{C}^{K\times 1}$, and $\mathbf s[n] \triangleq \left[s_1[n], \cdots, s_K[n]\right]^T\in \mathbb{C}^{K\times 1}$. It is not difficult to see that estimating $\boldsymbol \beta_m[i]$, $0\leq i \leq \mu$, is sufficient for the estimation of    both $\beta_{mkl}$ and $n_{kl}$ for the $KL$ paths. Note that \eqref{eq:rmn6} is equivalent to a MISO multi-path channel with $\mu+1$ taps, whose channel impulse response estimation has been extensively studied in the literature \cite{852}, \cite{853}. To this end, \eqref{eq:rmn6} is compactly written as \eqref{eq:rmBold} shown at the top of the page,
\begin{figure*}
\begin{align}\label{eq:rmBold}
\underbrace{\left[\begin{matrix}r_m[1]\\ r_m[2]\\ \vdots \\ r_m[T_2]
\end{matrix}\right]}_{\mathbf r_m}
=\sqrt{p_\tr} C\underbrace{\left[\begin{matrix} \mathbf s^T[1] & \mathbf 0 & \cdots & \mathbf 0  \\
\mathbf s^T[2] & \mathbf s^T[1] & \cdots & \mathbf 0 \\
\vdots & \vdots & \vdots & \vdots  \\
\mathbf s^T[T_2] & \mathbf s^T[T_2-1] & \cdots &  \mathbf s^T[T_2-\mu]\\
\end{matrix}
\right]}_{\mathbf S}
\underbrace{\left[\begin{matrix}\boldsymbol \beta_m[0] \\ \boldsymbol \beta_m[1] \\ \vdots \\ \boldsymbol \beta_m[\mu]
\end{matrix}\right]}_{\boldsymbol \beta_m} + \mathbf z_m,
\end{align}
\end{figure*}
where $\mathbf r_m\in \mathbb{C}^{T_2\times 1}$, $\mathbf S\in \mathbb{C}^{T_2\times K(\mu+1)}$, and $\boldsymbol \beta_m \in \mathbb{C}^{K(\mu+1)\times 1}$. Based on the received training signal $\mathbf r_m$ and the known pilots $\mathbf S$, the BS antenna $m$ can estimate $\boldsymbol \beta_m$ using the least square (LS) estimation, i.e.,
\begin{align}
\hat {\boldsymbol \beta}_m= \frac{1}{\sqrt{p_\tr} C}\left(\mathbf S^H \mathbf S\right)^{-1}\mathbf S^H \mathbf r_m.
\end{align}
In order to have a feasible estimation, the number of observations must be no smaller than the number of unknowns. Thus, we need $T_2\geq K(\mu+1)$.
Based on the estimation $\hat{\boldsymbol \beta}_m$, each BS antenna $m$ is associated with one path out of all the $KL$ paths. For $m\in \mathcal M_S$, let $(k_m^\star, i_m^\star)$ correspond to the strongest path arriving at BS antenna $m$, i.e.,
\begin{align}
(k_m^\star, i_m^\star) = \mathrm {arg} \ \underset{\substack{k=1,\cdots, K\\ i=0,\cdots, \mu}}{\max}   \big |\hat \beta_{mk}[i]\big|^2,
\end{align}
where $\hat{\beta}_{mk}[i]$, $k=1,\cdots, K$, $i=0, \cdots, \mu$, are the elements of the estimated tap gains in $\hat{\boldsymbol \beta}_m$. Then the desired path $(k_m, i_m)$ associated with BS antenna $m$ is set as
\begin{align}
(k_m, i_m)=
\begin{cases}
(k_m^\star, i_m^\star), & \text { if } {\substack{\big |\hat \beta_{mk_m^\star}[i_m^\star]\big|^2 \geq \\ \rho \sum_{(k,i)\neq (k_m^\star, i_m^\star)} \big | \hat \beta_{mk}[i]\big|^2}} \\
 \emptyset & \text{ otherwise},\label{eq:kmim}
\end{cases}
\end{align}
where $\rho\geq 0$ is a certain threshold. In other words, the BS antenna $m$ is associated with the strongest path $(k_m^\star, i_m^\star)$ if the ratio of the received power from this path and  that over all other paths is greater than  a threshold $\rho$; otherwise, no path is associated with this antenna. 
Note that in the subsequent data transmission phase, if BS antenna $m\in \mathcal M_S$ is synchronized to path $(k_m, i_m)$, then the signals received from all other paths will cause the detrimental ISI  or IUI. Intuitively, the received signal by BS antenna $m$ would have non-negligible contribution for signal detection only if its received strongest path dominates over all other paths, i.e., the first case of \eqref{eq:kmim} is true. Note that the path association rule in \eqref{eq:kmim} is also applicable when antenna $m$ receives no significant power from any of the $KL$ paths, in which case none of the path dominates and hence antenna $m$ will not be used for signal detection during the subsequent data transmission phase.

Note that with $(k_m, i_m)$ obtained, the corresponding path in terms of $(k_m, l_m)$ and the path delay $n_{k_ml_m}=i_m$ can be obtained accordingly.
Based on the above discussions, the minimum duration required for phase 2 is $\mu+K(\mu+1)$.

\subsection{Reduced MIMO Channel Estimation}
After phase 2 training, each of the selected BS antenna $m\in \mathcal M_S$ is associated with at most one desired signal path $(k_m, l_m)$ with its path delay $n_{k_ml_m}$ estimated. Recall that $\mathcal M_k\subset \mathcal M_S$ is defined as the subset of the BS antennas that are associated with MS $k$, i.e., $\mathcal M_k=\{m\in \mathcal M_S: k_m=k\}$. Then we have $\mathcal M_k \bigcap \mathcal M_{k'}=\emptyset$, $\forall k\neq k'$. The objective of phase 3 training is to estimate the effective MIMO channel for each MS $k$ with its associated BS antennas in $\mathcal M_k$. At the beginning of phase 3 training, a guard time of duration $\mu$ is inserted to avoid receiving the training symbols sent in phase 2, as illustrated in Fig.~\ref{F:ChannelEstimationFrameStructure}. After that, each MS $k$ sends {\it identical} training symbol $s$ using a set of analog beamforming vectors that span the whole $M_S$-dimensional space. Specifically, the $M_S$-dimensional training signal vector $\mathbf x_k[n]$ sent by MS $k$ is
\begin{align}
\mathbf x_k[n] = \sqrt{p_\tr}\mathbf f[n]s, \ n=1,\cdots, M_{\MS},
\end{align}
where $p_\tr$ is the training power, $\mathbf f[n]$ is the analog beamforming vector applied by the MSs at training duration $n$, and $M_\MS$ is the minimum number of training duration in phase 3 for the analog beamformers to span the whole $M_{\MS}$-dimensional space. Note that $\{\mathbf f[n]\}_{n=1}^{M_S}$ are chosen such that $\mathbf F\triangleq \big[ \mathbf f[1], \cdots \mathbf f[M_\MS] \big]\in \mathbb{C}^{M_\MS\times M_\MS}$ is a non-singular matrix.

For each MS $k$, the signals received by its associated BS antennas $m\in \mathcal M_k$ can be written as
\begin{align}
r_m&[n] = \sum_{k'=1}^K \sum_{l=1}^L \mathbf h_{mk'l}^H\mathbf x_{k'}[n-n_{k'l}]+ z_m[n]\notag \\
=&\underbrace{\sqrt{p_\tr}\mathbf h_{mkl_m}^H \mathbf f[n-n_{kl_m}]s}_{\text{desired signal}} + \underbrace{\sqrt{p_\tr}\sum_{l\neq l_m}^L \mathbf h_{mkl}^H \mathbf f[n-n_{kl}]s}_{\text{ISI}}\notag \\
 &\hspace{-3ex}+ \underbrace{\sqrt{p_\tr}\sum_{k'\neq k}^K \sum_{l=1}^L \mathbf h_{mk'l}^H \mathbf f[n-n_{k'l}]}_{\text{IUI}}+ z_m[n], \ n=1,\cdots,  M_\MS+\mu.
\end{align}
Note that $\mathbf f[n]=\mathbf 0$ for $n\leq 0$ or $n>M_\MS$.
As the BS antenna $m$ has the knowledge of path delay $n_{kl_m}$ of its associated path, it can apply the path delay compensation to $r_m[n]$ by letting $\bar r_m[n] \triangleq r_m[n+n_{kl_m}]$, which yields
\begin{align}
\bar r_m &[n] =
 \underbrace{\sqrt{p_\tr}\mathbf h_{mkl_m}^H \mathbf f[n]s}_{\text{desired signal}} + \underbrace{\sqrt{p_\tr}\sum_{l\neq l_m}^L \mathbf h_{mkl}^H \mathbf f[n-\Delta_{kl, kl_m}]s}_{\text{ISI}}\notag \\
&\hspace{-4ex} + \underbrace{\sqrt{p_\tr}\sum_{k'\neq k}^K \sum_{l=1}^L \mathbf h_{mk'l}^H \mathbf f[n-\Delta_{k'l, kl_m}]}_{\text{IUI}}+ z_m[n],  n=1,\cdots, M_\MS, \label{eq:rbarm}
\end{align}
where $\Delta_{k'l', kl}\triangleq n_{k'l'}-n_{kl}$ denotes the excessive path delay between path $(k',l')$ and $(k,l)$. By concatenating $\bar r_m[n]$ for all $n=1,\cdots, M_\MS$, \eqref{eq:rbarm} can be compactly written as
\begin{align}
\mathbf {\bar r}_m^H = \sqrt{p_\tr} \mathbf h_{mkl_m}^H \mathbf F s + \mathbf z^H_{m, \text{ISI}} + \mathbf z^H_{m,\text{IUI}}+ \mathbf z^H_m, \ m\in \mathcal M_k,\label{eq:barrmH}
\end{align}
where $\mathbf {\bar r}_m^H\triangleq \left[\bar r_m[1],\cdots, \bar r_m[M_\MS] \right]\in \mathbb{C}^{1\times M_\MS}$, and $\mathbf z^H_{m, \text{ISI}}, \mathbf z^H_{m,\text{IUI}}, \mathbf z_m^H \in \mathbb{C}^{1\times M_\MS}$ denote the concatenated ISI, IUI, and noise, respectively.

By further concatenating $\mathbf {\bar r}_m^H$ for all $m\in \mathcal M_k$ in \eqref{eq:barrmH}, we have
\begin{align}
\bar {\mathbf R}_k=\sqrt{p_\tr}\mathbf G_k \mathbf F s + \mathbf Z_{k,\text{ISI}} + \mathbf Z_{k,\text{IUI}} + \mathbf Z_k, \ k=1,\cdots, K, \label{eq:barRk}
\end{align}
where $\bar {\mathbf R}_k\in \mathbb{C}^{|\mathcal M_k|\times M_\MS}$ with rows given by $\bar{\mathbf r}_m^H$, $m\in \mathcal M_k$, is the effectively received training symbols by all the $|\mathcal M_k|$ BS antennas associated with MS $k$, $\mathbf G_k\in \mathbb{C}^{|\mathcal M_k|\times M_\MS}$ with rows given by $\mathbf h^H_{mkl_m}$ is the effective frequency-flat MIMO channel matrix from MS $k$ to its associated BS antennas after path delay compensation, and $\mathbf Z_{k,\text{ISI}}, \mathbf Z_{k,\text{IUI}}, \mathbf Z_k\in \mathbb{C}^{|\mathcal M_k|\times M_\MS}$ are the corresponding ISI, IUI, and noise matrix for MS $k$. By treating ISI and IUI as noise, the effective MIMO channel matrix $\mathbf G_k$ for MS $k$ can be estimated based on \eqref{eq:barRk} with the LS estimation, i.e.,
\begin{align}
\hat{\mathbf G}_k = \frac{1}{\sqrt{p_\tr}}\bar{\mathbf R}_k{\mathbf F}^{-1}s^*, \ k=1,\cdots, K. \label{eq:hatGk}
\end{align}
Note that the effective MIMO channel estimation in \eqref{eq:hatGk} is in general subject to the ISI and IUI contaminations. However, based on the path association rule in \eqref{eq:kmim}, for each BS antenna that is associated with one MS, the ISI and IUI is dominated by the desired signal path, and hence their  detrimental effect on channel estimation can be properly mitigated via choosing an appropriate threshold $\rho$ in \eqref{eq:kmim}. In particular, in the favorable scenario when all paths are well separated at the BS such that each BS antenna $m$ receives non-negligible power from at most one signal path, the ISI and IUI for channel estimation in \eqref{eq:hatGk} is negligible even for $\rho=0$.

After obtaining the effective MIMO channel estimation $\hat {\mathbf G}_k$ for each MS $k$, the BS obtains the optimized transmit and receive beamforming vectors $( \hat{\mathbf v}_k^\star, \hat  {\mathbf u}_k^\star)$ based on \eqref{eq:uk1vk1} with $\mathbf G_k$ replaced by $\hat {\mathbf G}_k$. It then sends back the index of the analog beamforming vector $\hat{\mathbf v}_k^\star\in \mathcal V$ to each MS $k$. The total number of required feedback bits is $K\log_2(N_\CB)$.

Based on the previous discussions, the total time overhead required for phase 3 is $M_\MS+2\mu$. Thus, the total time required for the proposed channel estimation scheme is
$T=\lceil \frac{M_{\BS}}{M_{\RF}}\rceil + M_\MS + 4\mu +K(\mu+1)$, which is much smaller than $T'$ as required by the brute force channel impulse response estimation. For the example given at the beginning of this section, we have $T=503\ll 128000$. Besides, as compared to the channel coherence time $T_c=50000$, the  time overhead of the proposed channel estimation scheme is $1.0\%$ and thus negligible.

The frame structure of the proposed channel estimation scheme is illustrated in Fig.~\ref{F:ChannelEstimationFrameStructure}.

\section{Simulation Results}\label{sec:simulation}
In this section, simulation results are provided to compare the performance of the proposed lens-based system with that based on the conventional antenna arrays. For both systems, we assume that each MS is equipped with the conventional UPA with $M_S=16$ elements, whereas the BS either has a FD lens array (proposed), or a UPA (benchmark) with adjacent elements separated by half wavelength. We assume that both the lens array and the UPA at the BS have the same effective aperture $\D_y\times \D_z=10 \times 10$, and the lens array is designed to have the maximum coverage angles $\Phi=\Theta=180^\circ$. Thus, the total number of BS antennas for UPA is $400$ and that for the lens array can be obtained as $317$. We assume that the system operates at $28$ GHz and the mmWave channel for each MS has $L=3$ paths, with the azimuth and elevation angles independently and uniformly distributed in $[-60^\circ, 60^\circ]$. Furthermore, the path delays are uniformly distributed in $[0, T_m]$, with $T_m=100$ns denoting the maximum path delay. We assume that all MSs are $100$m away from the BS, and the path loss and power division among different multi-paths are generated based on the model developed in \cite{565}. We further assume that the total available bandwidth is $B=500$MHz. As a result, we have $\mu=BT_m=50\gg 1$, so that the system is wide-band and frequency-selective in general. For the benchmark UPA system, we assume that perfect CSI is available at the BS and MSs and MIMO-OFDM transmission is adopted, with $N=512$ sub-carriers and $\mu=50$ cyclic prefix (CP) symbols. To cater for the limited number of RF chains at the BS side, the approximate Gram-Schmidt based hybrid analog/digital signal processing scheme for MIMO-OFDM proposed in \cite{825} is applied for the UPA system. On the other hand, for the lens system, we consider both cases with perfect CSI as well as the estimated CSI as proposed in Section~\ref{sec:channelEstimation}, where the threshold $\rho$ in \eqref{eq:kmim} for channel estimation is set to zero. After the power-based antenna selection in the lens system, we employ the proposed MRC-based (for both perfect and estimated CSI) and MMSE-based (for perfect CSI only) PDMA schemes, which only require the low-complexity SC transmission and path delay compensation at the BS. At the MS side, we assume that both the UPA and lens systems apply analog beamforming with the beamsteering codebook of size $N_\CB=256$, which is obtained by uniformly quantizing the azimuth and elevation angles. Note that to account for the training overhead of the proposed channel estimation scheme, we assume that the channels are quasi-static that remain unchanged for $T_c=0.1$ms, or equivalently about $50000$ symbols for SC transmission with bandwidth $B=500$MHz. Thus, the training overhead can be calculated based on Section~\ref{sec:channelEstimation} for different number of RF chains, $M_\RF$.

\begin{figure}
\centering
\includegraphics[scale=0.28]{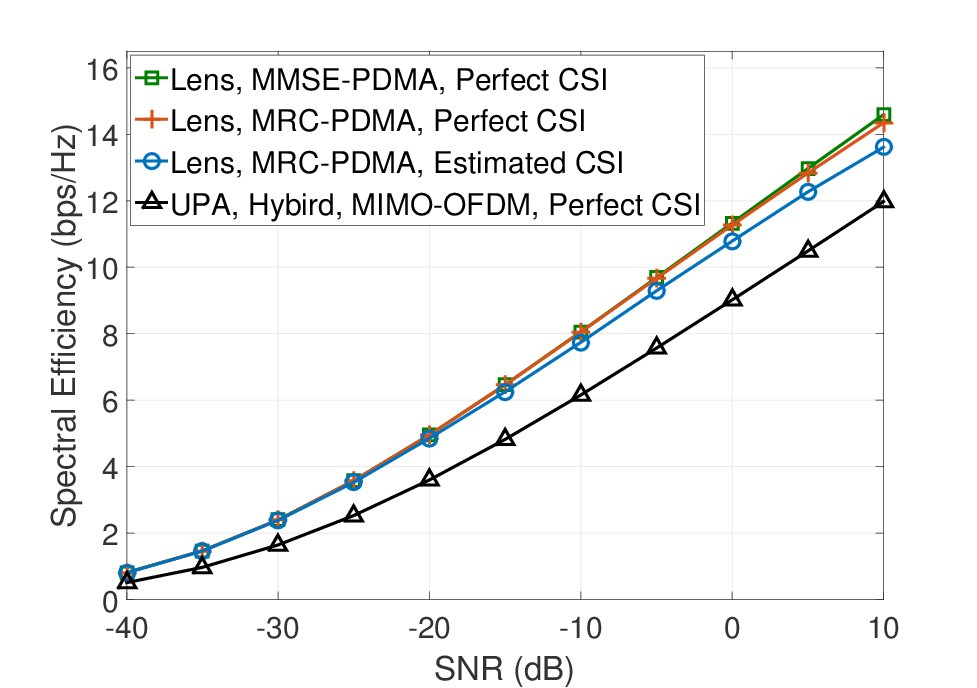}
\caption{Spectrum efficiency versus SNR for single-user mmWave systems.}\label{F:RateVsSNRSingleuser3RFChain}
\end{figure}

\subsection{Single-User System}
First, we consider the special case of single-user system with $K=1$, for which the hybrid processing scheme  proposed in \cite{825} can be directly applied. By assuming that the BS has $M_\RF=3$ RF chains, Fig.~\ref{F:RateVsSNRSingleuser3RFChain} shows the spectrum efficiency versus the SNR of the data transmission phase for various schemes. It is observed that for the lens system with perfect CSI, the proposed MRC- and MMSE-based schemes achieve almost identical performance. This is expected since in the single-user setup with a total of three paths only, all multi-path signals are well separated at the BS almost surely and thus the IPI (or equivalently ISI in this case) vanishes. Furthermore, by setting the training SNR as 10dB, it is found that the performance of the MRC-based scheme with the estimated CSI is very close to that based on perfect CSI, which shows the efficacy of the  channel estimation scheme proposed in Section~\ref{sec:channelEstimation}. Note that for the setup under consideration, the required training length for each channel coherent block can be calculated as 373, which is only about 0.75\% of the channel block length and hence is negligible. Moreover, it is noted from Fig.~\ref{F:RateVsSNRSingleuser3RFChain} that even with estimated CSI, the proposed lens system achieves higher spectrum efficiency than the benchmark UPA system based on perfect CSI, which requires the more sophisticated MIMO-OFDM and hybrid signal processing schemes. The performance gain is mainly attributed to the saving of the CP overhead with the proposed lens-based design, the efficacy of the proposed channel estimation scheme with almost negligible training overhead, as well as the effective  ISI mitigation and coherent signal combining after path delay compensation at the BS.

\begin{figure}
\centering
\includegraphics[scale=0.28]{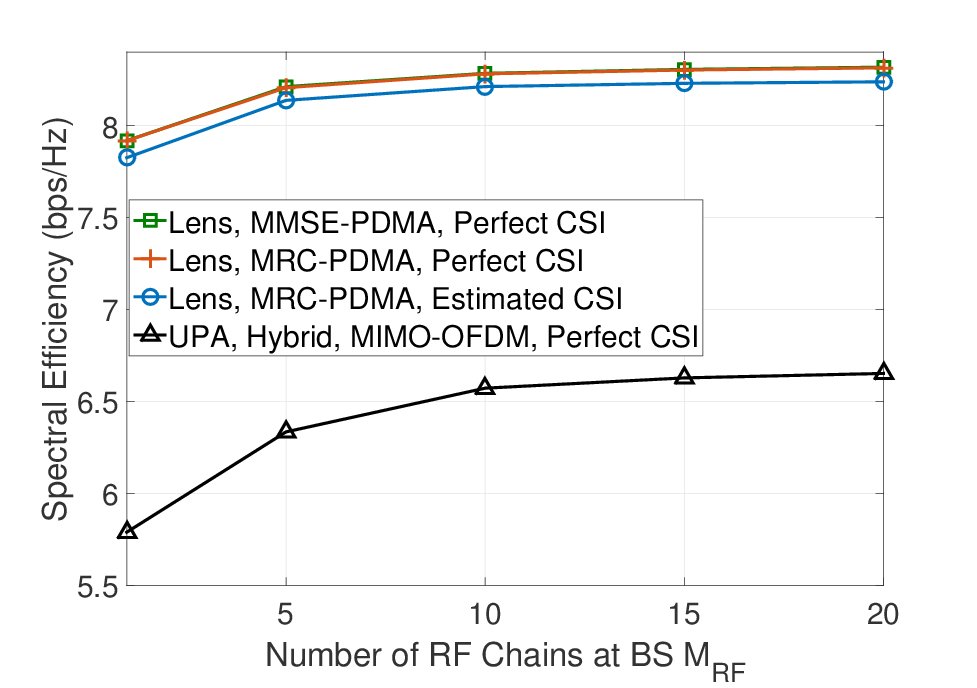}
\caption{Spectrum efficiency versus the number of RF chains $M_\RF$ at the BS for single-user mmWave systems.}\label{F:RateVsRFChainSingleuserSNRMinus10}
\end{figure}

Fig.~\ref{F:RateVsRFChainSingleuserSNRMinus10} shows the achievable spectrum efficiency versus the number of RF chains $M_\RF$ at the BS, where the SNR for the data transmission phase is set as $-10$ dB. It is observed that for all the schemes, the performance in general enhances as more RF chains are available at the BS, but only marginal improvement is observed as $M_\RF$ exceeds $10$. In particular, for all the schemes under consideration, by equipping the BS with only $5$ RF chains is able to achieve over $99\%$ of the spectrum efficiency as compared to the full RF chain case. This is expected since as $M_\RF$ increases, the system performance will be eventually constrained by the limited number of signal paths of the mmWave channels.



\begin{figure}
\centering
\includegraphics[scale=0.28]{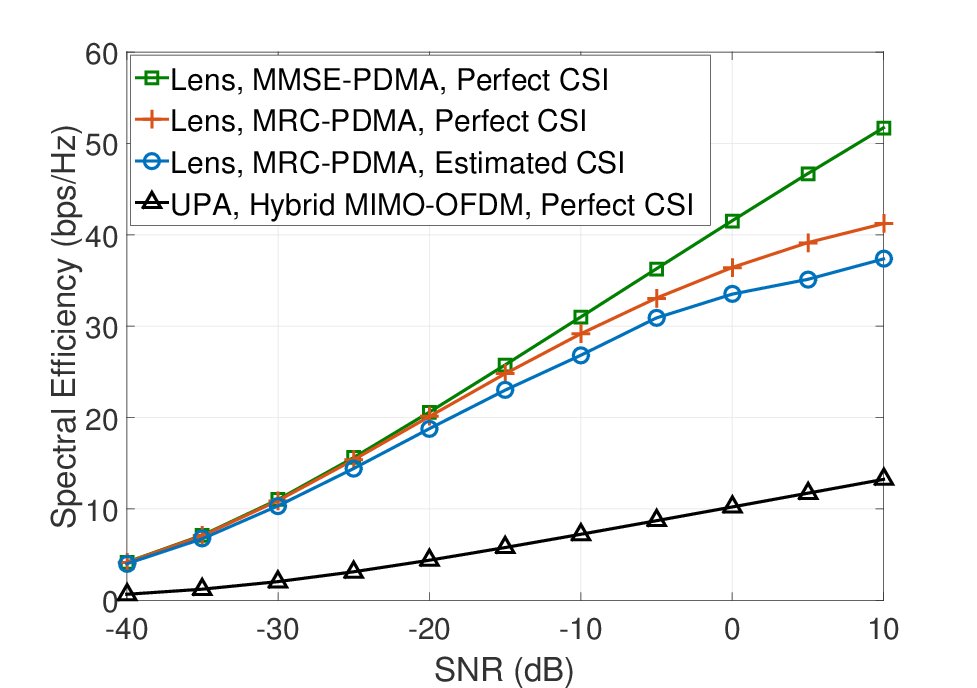}
\caption{Spectrum efficiency versus SNR for multi-user mmWave systems.}\label{F:RateVsSNR5user10RFChain}
\end{figure}

\subsection{Multi-User System}
Next, we consider the multi-user system with $K=5$ MSs. Since, to the best of the authors' knowledge, no hybrid processing designs were reported for the multi-user wide-band MIMO-OFDM systems, the benchmark scheme is simply chosen to be the hybrid scheme proposed in \cite{825} for single-user MIMO-OFDM system together with time division multiple access (TDMA), i.e., each of the $K$ MSs is served by the BS for $1/K$ of the time. Fig.~\ref{F:RateVsSNR5user10RFChain} shows the achievable spectrum efficiency for different schemes by assuming the BS has $M_\RF=10$ RF chains. It is observed that for the lens MIMO system, the MMSE- and MRC-based PDMA schemes have almost identical performance in the low-SNR regime, whereas the performance gap increases as the SNR increases. This is expected since for multi-user setup with more signal paths in total, the IPI becomes stronger as the SNR increases. Furthermore, by setting the training SNR as $20$ dB, it is found that the proposed channel estimation scheme results in comparable performance as the perfect CSI case, which shows its effectiveness for multi-user system as well. Fig.~\ref{F:RateVsSNR5user10RFChain} also shows that the proposed lens system significantly outperforms the UPA system, mainly due to the more efficient multiple-access scheme (PDMA) compared to TDMA. However, even by ignoring the time division loss of TDMA (via multiplying the sum-rate by a factor of $K$) and thus obtaining a (loose) performance upper bound for the UPA-based hybrid system, the proposed lens system still performs comparably well for all SNRs.

\begin{figure}
\centering
\includegraphics[scale=0.28]{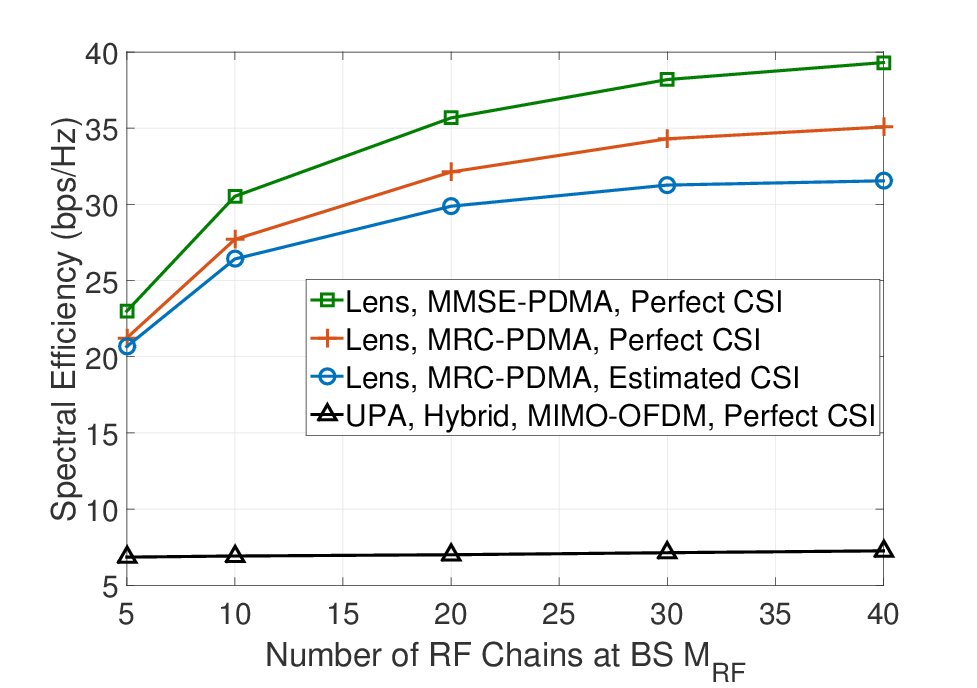}
\caption{Spectrum efficiency versus the number of RF chains $M_\RF$ at the BS for multi-user mmWave systems.}\label{F:RateVsRFChain5userSNRMinus10}
\end{figure}

  Fig.~\ref{F:RateVsRFChain5userSNRMinus10} shows the achievable spectrum efficiency for the multi-user system versus the number of RF chains $M_\RF$ at the BS, where the SNR for data transmission is set as $-10$ dB. Similar to the single-user case, the spectrum efficiency of the lens-based scheme for multi-user systems increases  with $M_\RF$. In particular,  with only $M_\RF=20$ RF chains at the BS, about $90\%$ of the spectrum efficiency achievable by the full RF chain system can be attained. Fig.~\ref{F:RateVsRFChain5userSNRMinus10} also shows that as $M_\RF$ increases, there is no evident performance improvement for the benchmark UPA system, since at each instance the BS only serves one MS with a single data stream due to the use of TDMA.

\section{Conclusion}\label{sec:Conclu}
This paper studies the uplink multi-user mmWave MIMO communication with single-sided FD lens antenna array at the BS and conventional UPA at each MS. Under limited RF chains at the BS and one single RF chain at each MS, we first propose an efficient PDMA scheme, by which each MS essentially communicates with the BS via different channel paths with the low-complexity SC transmission and  path delay compensation at the BS. For general scenarios with insufficiently separated AoAs, analog beamforming at the MSs is jointly designed with  digital combining at the BS based on MRC and MMSE, respectively. Furthermore, we propose an efficient channel estimation scheme tailored for the MRC-based PDMA scheme, which incurs almost negligible training overhead in practical mmWave systems and yields comparable performance as the case with perfect CSI. Numerical results are provided to show the sum-rate gain of the proposed design over the benchmark UPA systems based on MIMO-OFDM and hybrid analog/digital signal processing at the BS, which in general require higher signal processing complexity and hardware/energy costs than the proposed system.

\appendices
\section{Proof of Lemma~\ref{lemma:response}}\label{A:proof}
The proof of the array response in \eqref{eq:responseFDLens} extends that in Appendix~A of \cite{823} by considering the signals' elevation angles. We will assume that the lens array is used for signal reception, and the proof of transmit array response can be obtained similarly due to reciprocity. Let $D_y \times D_z$ denote the {\it physical dimension} of the EM lens, and $\Phi(y,z)$ with $(y,z)\in \left[-D_y/2, D_y/2\right]\times \left[-D_z/2, D_z/2\right]$ denote its phase shift profile, which represents the phase delay provided by the spatial phase shifters (SPS) at each point $(0, y, z)$ on the lens's aperture. Further denote by $B_0$ with coordinate $(F,0,0)$ the focal point of the lens for normal incident plane waves, where $F$ is the focal length. To ensure constructive superpositions at $B_0$ for all rays with normal incidence, $\Phi(y,z)$ must be designed to be \cite{823}
\begin{align}
\Phi(y,z)& =\Phi_0-k_0\sqrt{F^2+y^2+z^2}, \notag \\
 &\forall (y, z) \in \left[-\frac{D_y}{2}, \frac{D_y}{2}\right]\times \left[-\frac{D_z}{2}, \frac{D_z}{2}\right], \label{eq:Phiyz}
\end{align}
where $\Phi_0$ denotes the common phase shift from the lens's input aperture to the focal point $B_0$, and $k_0=2\pi/\lambda$ is the wave number corresponding to the signal wavelength $\lambda$. With $\Phi(y,z)$ designed as in \eqref{eq:Phiyz}, the resulting phase delay from the lens's input aperture $(0,y,z)$ to antenna $m$ with coordinate $B_m(F\cos\theta_m\cos\phi_m, F\cos\theta_m\sin\phi_m, F\sin\theta_m)$ can be expressed as
\begin{align}
\psi_m&(y,z)=\Phi(y,z)+k_0d_m(y,z)\notag \\
&=\Phi_0-k_0\sqrt{F^2+y^2+z^2}\notag \\
&+k_0 \sqrt{F^2+y^2+z^2-2yF\cos\theta_m\sin\phi_m-2zF\sin\theta_m}\notag\\
&\approx \Phi_0-k_0y\cos\theta_m\sin\phi_m-k_0z\sin\theta_m, \label{eq:Taylor}
\end{align}
where $d_m(y,z)$ denotes the distance from the point $(0,y,z)$ on the lens to antenna $m$, and \eqref{eq:Taylor} follows from the first-order Taylor approximation with the assumption $F\gg D_y, D_z$.

Denote by $s(y,z)$ the arriving signal at point $(y,z,0)$ of the lens's input aperture. Due to the linear superposition principle, the resultant signal at antenna $m$ can then be expressed as
\begin{align}
r_m(\theta, \phi)=\int_{-D_z/2}^{D_z/2}\int_{-D_y/2}^{D_y/2} s(y,z)e^{-j\psi_m(y,z)}d_yd_z. \label{eq:rm}
\end{align}

For a uniform incident plane wave with elevation AoA $\theta$ and azimuth AoA $\phi$, the arriving $s(y,z)$ on the lens aperture can be expressed as
\begin{align}
s(y,z)=\frac{1}{\sqrt{\beta}}x_0 e^{-jk_0(y\cos\theta\sin\phi+z\sin\theta)}, \label{eq:syz}
\end{align}
where $x_0$ is the input signal arriving at the reference point (chosen as the lens center) of the lens, and $\beta\triangleq \lambda^2 D_yD_z$ is a normalization factor to ensure that the total power captured by the lens is proportional to its effective aperture $\D_y\D_z$. By substituting \eqref{eq:syz} and \eqref{eq:Taylor} into \eqref{eq:rm}, $r_m(\theta, \phi)$ can be written as \eqref{eq:rm2} shown at the top of the next page,
\begin{figure*}
\begin{align}
r_m(\theta, \phi)\approx & x_0 e^{-j\Phi_0}\sqrt{\beta}\int_{-D_z/2}^{D_z/2}e^{-jk_0z(\sin \theta - \sin\theta_m)}dz \times \int_{-D_y/2}^{D_y/2}e^{-jk_0y(\cos\theta\sin\theta-\cos\theta_m\sin\theta_m)}dy \notag \\
=&x_0e^{-j\Phi_0}\sqrt{\D_y\D_z} \sinc \left(\D_z\left(\sin\theta_m-\sin\theta\right) \right) \sinc \left(\D_y\left(\cos\theta_m\cos\phi_m-\cos\theta\cos\phi\right) \right), \label{eq:rm2}
\end{align}
\end{figure*}
where we have used the identity $\D_y\triangleq D_y/\lambda$ and $\D_z\triangleq D_z/\lambda$.
By substituting \eqref{eq:me} and \eqref{eq:ma} into \eqref{eq:rm2} and with the definition $a_m(\theta, \phi)\triangleq r_m(\theta,\phi)/x_0$, the array response in \eqref{eq:responseFDLens} can be obtained.

This completes the proof of Lemma~\ref{lemma:response}.

\bibliographystyle{IEEEtran}
\bibliography{IEEEabrv,IEEEfull}

\end{document}